%% file: reverso.tex
\documentclass[table,sigconf]{acmart}
\setcopyright{none}
\renewcommand\footnotetextcopyrightpermission[1]{}

\usepackage{balance}

\usepackage{xcolor}

\usepackage{amsmath,amsfonts}
\usepackage{algorithmic}
\usepackage{bytefield}
\usepackage{textcomp}
\usepackage{bmpsize}
\usepackage{graphicx}
\usepackage{pgf}
\usepackage{standalone}
\usepackage{tikz}
\usepackage{adjustbox}
\usetikzlibrary{calc,positioning,shapes,decorations.pathreplacing, arrows.meta}
\usepackage{paralist}
\usepackage{listings, listings-rust}
\usepackage{booktabs}
\usepackage[detect-all]{siunitx}
\usepackage{subcaption}
\usepackage{tabularx}
\usepackage{hyperref}
\def\BibTeX{{\rm B\kern-.05em{\sc i\kern-.025em b}\kern-.08em
    T\kern-.1667em\lower.7ex\hbox{E}\kern-.125emX}}

\newcommand{\reverso}{\texttt{quiceh}}
\tikzset{
short/.style={draw,rectangle,text height=4pt,text depth=13pt,
  text width=7pt,align=center,fill=gray!30},
long/.style={short,text width=1.5cm}
}

\def\labelnode#1#2#3#4{
  \node[long, auto, text width=#3, right=of #1, label=center:#4] (#2) {}}

\def\chunknode#1#2#3#4#5{
  \node[rectangle, fill=gray!30, draw, text height=#3, text width=#4, right=of #1,
  align=center]
  (#2) {#5}}

\def\encnode#1#2#3{
  \node[right=of #1, thick, draw=red, rectangle, text height=4pt, text depth=15 pt,
  minimum width=#3] (#2inner) {};
  \node[right=of #2inner, thick, draw=red, rectangle, text height=4pt, text depth=15pt, text width=7pt] (#2) {\rotatebox{270}{Tag}}}

\usepackage[breakable, theorems, skins]{tcolorbox}
\tcbset{enhanced}

\DeclareRobustCommand{\mybox}[2][gray!20]{%
\begin{tcolorbox}[   
        breakable,
        left=0pt,
        right=0pt,
        top=0pt,
        bottom=0pt,
        colback=#1,
        colframe=#1,
        width=\dimexpr\columnwidth\relax,
        enlarge left by=0mm,
        boxsep=5pt,
        arc=0pt,outer arc=0pt,
        ]
        #2
\end{tcolorbox}
}
\begin{document}


\title{Contiguous Zero-Copy for Encrypted Transport Protocols}

\author{Florentin Rochet}
\affiliation{
	\institution{UNamur, Belgium}
}
\email{florentin.rochet@unamur.be}

\begin{abstract}
  We propose in this paper to revisit the design of existing encrypted
  transport protocols to improve their efficiency. 
  We call the methodology
  ``Reverso'' from reversing the order of field elements within a protocol
  specification. We detail how such a benign-looking change within the
  specifications may unlock contiguous zero-copy for encrypted protocols during
  data transport. To demonstrate our findings, we release \texttt{quiceh}, a
  QUIC implementation of QUIC VReverso, an extension of the QUIC V1 standard
  (RFC9000).  Our methodology applied to the QUIC protocol reports $\approx
  30\%$ of CPU efficiency improvement for processing packets at no added cost
  on the sender side and without relaxing any security guarantee from QUIC V1.
  We also implement a fork of Cloudflare's HTTP/3 module and client/server
  demonstrator using \texttt{quiceh} and show our optimizations to directly
  transfer to HTTP/3 as well, resulting in our new HTTP/3 to be $\approx 38\%$
  more efficient than the baseline implementation using QUIC V1. We argue that
  Reverso applies to any modern encrypted protocol and its implementations and
  that similar efficiency improvement can also be unlocked for them,
  independently of the layer in which they operate.  Indeed, this research
  shows that the ability to implement contiguous zero-copy on the receiver side
  inherently depends on the specified encrypted protocol wire image, and that
  we may need to reverse how we are used to write them.
\end{abstract}



\maketitle

\input{body}

\section*{Acknowledgments}
We would like to express our sincere gratitude to Lionel Goffaux, Laurent
Schumacher, Kazuho Oku and Gilles Perrouin for providing feedbacks to earlier
versions of this manuscript. This research
also benefited from fruitful discussions with Tobias Pulls, Gaëtan Cassiers and
Nick Banks. We thanks the reviewers from NSDI, IEEE Euro S\&P and ACM SIGCOMM CCR
for meaningful feedbacks to earlier versions of this document. Any error or
omission is entirely the author's own responsibility.
This research was partially funded by the CyberExcellence project of the Public
Service of Wallonia (SPW Recherche), convention No. 2110186.
\bibliographystyle{ACM-Reference-Format}
\bibliography{reverso}

\appendix
\input{1_appendix}
\end{document}

%% file: body.tex
\section{Introduction}
\input{1_introduction}

\section{Background}
\input{2_background}

\section{Protocol Reverso}
\input{3_protocol_reverso}
\section{Reverso for QUIC}
\input{4_reverso_for_quic}
\section{A tour of QUIC implementations}
\input{5_implem}
\section{Efficiency Evaluation}
\input{6_perf_eval}
\section{Security and Safety of QUIC VReverso}
\input{7_sec.tex}

\section{Related Works}
\input{7_relatedwork}

\section{Conclusion}
\input{8_ccl}

%% file: 1_introduction.tex

Encrypted transport protocols evolved with different background contexts but
share similarities in design principles. Tor~\cite{tor}, for example, was
designed as a stream-based multihop TCP/IP overlay in the early 2000's,
encrypts 512-bytes sized packets called ``cells'' mixing routing information,
end-to-end transport control and end-to-end data. More recently,
QUIC~\cite{10.1145/3098822.3098842} and TCPLS~\cite{tcpls-conext} are designed
as general-purpose transport protocols rooted in being extensible, on offering
a 1-RTT handshake, and on being resilient to middlebox interference throughout
their evolution. Middlebox resilience is also a shared property of Tor, which
comes from fully encrypting control information and enforcing exact sizes,
preventing middleboxes from creating protocol ossification~\cite{180723} complexifying protocol evolution.  Other
protocol designs, such as TCP, DCCP~\cite{kohler2006designing} or
SCTP~\cite{stewart2001sctp,stewart2022rfc} apply encryption following an
encapsulation approach without changing much of their initial design.

Cryptography imposed itself as a fundamental requirement for transporting
information over the Internet. Applying cryptography is a fundamental paradigm
change in Internet communication that is ongoing, since the widespread
availability of cryptography acceleration in
hardware~\cite{akdemir2010breakthrough}. We argue that it should involve
questioning protocol wire format designs. We show that a part of the cost of
using cryptography within an established session in modern encrypted transport
protocols is caused by a conceptual misalignment with the wire format, preventing
implementers from efficiently manipulating existing interfaces to symmetric
cryptographic primitives, especially on the receiving side.

In this paper, we identify this misalignment and explain how to adapt the
protocol wire format using two key design principles to avoid protocol-induced data
fragmentation, causing memory copy overheads on the receiver's implementation.
Our protocol design principles permit the implementation of a contiguous zero-copy
abstraction on the receiver otherwise impossible to achieve in encrypted transports without
touching the established cryptography interface of symmetric encryption.  We
argue that modifying the current interface to symmetric encryption to obtain a
similar achievement could be possible but would be against the current state of
knowledge from related side-channel security literature promoting atomic
interfaces (i.e., output the whole decryption at once or an error).
Furthermore, our suggestions may apply to any future transport protocol
involving mixed encryption of control and data.

The structure of the paper is as follows: Section~\ref{sec:background}
introduces necessary background in symmetric encryption and the QUIC protocol,
Section~\ref{sec:reverso} introduces our proposal and how it generically
applies to protocols mixing encrypted control and upper-layer data, and a
discussion of alternative approaches.

To demonstrate our findings, we choose the QUIC protocol and discuss necessary
changes to the QUIC wire image in Section~\ref{sec:quic_reverso}.  We pursue in
Section~\ref{sec:implem} with a tour of some of the existing efficient QUIC v1
implementations written in C/C++ or Rust to discuss how QUIC specifications are
driving software engineering choices and how our findings would help with
existing trade-offs. We give insights about the expected efficiency improvement of
Reverso for these implementations, depending on their architecture and API
choices.

We then introduce \reverso, our portable QUIC implementation implementing
VReverso forked from Cloudflare's QUIC implementation. Our efficiency
benchmarks in Section~\ref{sec:bench} report $\approx 30\%$ efficiency increase
for processing QUIC packets compared to the baseline portable QUIC v1
architecture.  We also implement an HTTP/3 client and server using QUIC
VReverso, and demonstrate that our findings directly translate to HTTP/3 by
propagating \reverso's API to HTTP/3's API supporting applications to process
HTTP/3 frame content of arbitrary length in contiguous zero-copy. Eventually,
Section~\ref{sec:sec} discusses specific security
considerations for VReverso.

%% file: 2_background.tex
\label{sec:background}

\begin{figure*}[t]
  \centering
  \begin{tikzpicture}[node distance=-\pgflinewidth]

    \node[] (a) {};
    \chunknode{a}{b}{5pt}{1.5cm}{UDP Header};
    \labelnode{b}{c}{1cm}{Flags};
    \chunknode{c}{d}{5pt}{1.5cm}{Dest Conn ID};
    \chunknode{d}{e}{5pt}{1.5cm}{Packet Number};
    \chunknode{e}{f}{5pt}{1.5cm}{Control Frame};
    \chunknode{f}{g}{5pt}{1.5cm}{Control Frame};
    \chunknode{g}{h}{5pt}{3.5cm}{Stream\\Frame};
    \chunknode{h}{i}{5pt}{1.5cm}{Control Frame};
    \encnode{e}{i}{8.9cm};

    \draw (b.south east) -- +(0,-0.8cm);
    \draw (f.south west) -- +(0,-0.8cm);
    \draw[<->] ( $ (b.south east) +(0,-0.4cm) $ ) -- node[fill=white] {QUIC Short Header} ( $ (f.south west) +(0,-0.4cm) $ );
    \draw (i.south east) -- +(0,-0.8cm);
    \draw[<->] ( $ (e.south east) +(0,-0.4cm) $ ) -- node[fill=white] {QUIC Payload} ( $ (i.south east) +(0,-0.4cm) $ );

    \draw (e.north east) -- +(0,0.8cm);
    \draw (i.north west) -- +(0,0.8cm);
    \draw (c.north west) -- +(0,0.8cm);
    \draw[<->] ( $ (e.north east) +(0,0.4cm) $ ) -- node[fill=white] {AEAD Encrypted} ( $ (i.north west) +(0,0.4cm) $ );
    \draw[<->] ( $ (c.north west) +(0,0.4cm) $ ) -- node[fill=white] {Associated Data} ( $ (e.north east) +(0,0.4cm) $ );
    \draw[<->] ( $ (e.north west) +(0,0.8cm) $ ) -- ( $ (e.north east) +(0,0.8cm) $ );
    \draw[<->] ( $ (c.north west) +(0.4cm,0.8cm) $ ) -- ( $ (c.north east) +(0,0.8cm) $ );
    \draw[decorate,decoration={brace,raise=2pt}] ( $ (c.north west) +(0.8cm,0.82cm) $ ) -- node[above=6pt] {One-time pad Encrypted} ( $ (e.north east) +(-1cm, 0.82cm) $);
  \end{tikzpicture}
  \caption{Typical QUIC short header packet, exchanged after the session is established}
  \label{fig:quicpacket}
\end{figure*}
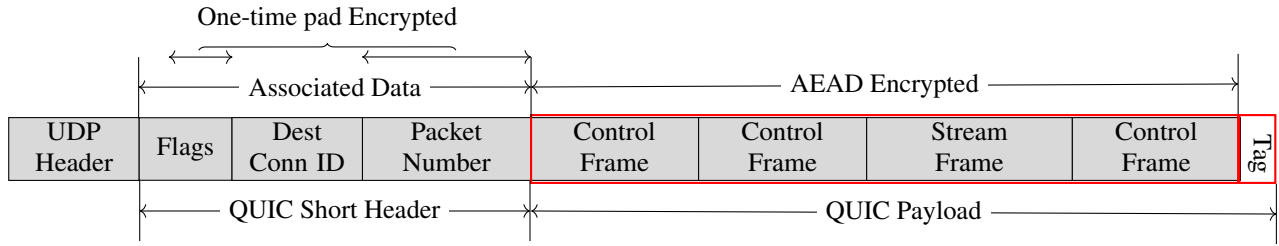

\subsection{AEAD ciphers with atomic interface}

AEAD~\cite{rogaway2002authenticated} stands for Authenticated Encryption with
Associated Data and regroups encryption schemes built to ensure data
confidentiality and data + associated data authenticity.  AEAD encryption
outputs the encrypted form of the input data and a tag computed from both the
data and the associated data and is usually attached at the end of the encrypted payload.
Security notions of such ciphers typically include semantic security under
chosen-plaintext attacks and ciphertext integrity thanks to the strong
unforgeability of the tag using an Encrypt-then-MAC composition
paradigm~\cite{bellare2008authenticated}. The decryption process verifies the
tag's validity by recomputing it using the encrypted data and the associated
data, then compares the result to the received tag. If those match, the
primitive outputs a decryption of the data.  If it does not
match, the primitive generates an error and decryption halts.

RFC5116~\cite{rfc5116} is the proposed standard
for interfacing AEADs with any software. In particular, for the decryption
operation, the RFC and related literature in protocol side-channel
attacks~\cite{albrecht2009plaintext, bernstein2012security, fischlin2015data}
suggest implementers not outputting any cleartext as long as
the whole ciphertext and associated data have not been verified. In
consequence, typical cryptography libraries with AEAD APIs are atomic; i.e., it
requires the operation to be done at once and explicitly prevents partial decryption. 


\subsection{QUIC}
\label{sec:backgroundquic}

QUIC~\cite{10.1145/3098822.3098842} is a transport protocol recently
standardized as RFC9000~\cite{rfc9000} and supports reliable, confidential, and
authenticated stream-based exchange of data over the Internet using an AEAD
cipher to encrypt both Application Data and QUIC control information. After the
handshake, whose goal is to derive an unpredictable and MITM-resistant
symmetric key efficiently, endpoints typically exchange UDP-encapsulated QUIC
packets, as depicted on Figure~\ref{fig:quicpacket}, containing a
\emph{protected} QUIC short header~\cite{rfc9001} and the authenticated and
confidential payload mixing control information and potential Application-level
data. The entire cleartext packet header is the AEAD associated data.

\noindent
\textbf{Two-levels encryption.} QUIC has a two-levels encryption/decryption
procedure (Figure~\ref{fig:quicpacket}): the payload is encrypted using an
AEAD, and part of the QUIC header is encrypted using a XOR with a key different from the AEAD key and obtained from a
per-packet key-derivation procedure detailed in RFC9001~\cite{rfc9001}, Section
"Header Protection".
The authenticity of the encrypted information within the header is
verified by using the cleartext header (after header
decryption) as associated data to the AEAD's decryption procedure of the payload.
If such decryption fails, then it means either the payload or the header has
been altered on the wire. Any usage of the decrypted QUIC header before payload
decryption occurs must be realized with caution and may leak information about
the encrypted header's content, as the information is not yet authenticated.
Typical QUIC V1 implementations do not make use of the decrypted QUIC header
before the AEAD's decryption succeeds. QUIC VReverso implementations may have
to perform a memory allocation after the header is decrypted but before the
header is authenticated, and there are safety and security considerations with
such an operation. We discuss them in Section~\ref{sec:sec}.

\noindent 
\textbf{QUIC features.}
QUIC involves features such as connection migration, flexible congestion
control, streams, and extensions, such as unreliable
datagrams~\cite{pauly2018unreliable} or
multipath~\cite{de2017multipath,quic-multipath}. Each of these features have
their own set of frames potentially mixed together in arbitrary order within a same encrypted
packet. QUIC is defined to be the underlying transport protocol for
HTTP/3~\cite{rfc9114} with a semantic mapping on both protocols' stream
abstractions.

%% file: 3_protocol_reverso.tex
\label{sec:reverso}
\subsection{The State of Encrypted Transport Protocols}

As of today, we integrate symmetric encryption into protocol implementations with a layering
approach: we have independence between symmetric cryptography usage and the protocol design. That is,
we encrypt/decrypt blobs of data without considering how their structure may
help packet processing efficiency involving encryption or decryption. In some cases, we
may make this choice to maintain backward compatibility or lower friction to deploy
encryption with established implementations. For other protocols that cannot
function without encryption, such as QUIC, Tor, or SSH, questioning the established
design practice offers opportunities.

To understand these opportunities, we first need to generalize encrypted
transports. Packets of encrypted transport protocols mixing encrypted control
and encrypted data may be characterized as follows:

\begin{equation}
    pkt := H_1 \parallel \text{ENC}_i(H_2) \parallel \text{ENC}_j(P_C)
\end{equation}

Where:
\begin{inparaenum}
    \item $\parallel$ denotes concatenation.
    \item $H_1$ is the cleartext transport header.
    \item $\text{ENC}_i(H_2)$ represents an encrypted header of the secure transport
      protocol using encryption scheme $i$.
    \item $\text{ENC}_j(P_C)$ contains the encrypted payload mixed with
      protocol-specific controls, potentially using a different encryption
      scheme $\text{ENC}_j \neq \text{ENC}_i$ and a different set of keys.
\end{inparaenum}

The payload $P_C$ may exhibit a structure combining both an arbitrary number of
upper layer data fragments and optional control information for the transport
protocol:

\begin{align*}
  &P_C ::= \text{<chunk>}\parallel \text{<chunk>}\parallel \ldots \parallel \text{<chunk>}\\
  &\text{<chunk>} ::= \left(\text{<header>}\parallel \text{<data>}\right) | \text{<header>}\\
  &\text{<header>} ::= \text{<field>}\parallel \ldots \parallel \text{<field>}\\
  &\text{<field>} ::= \text{<u8>} | \text{<u16>} | \text{<u32>} | \text{<u64>} | \text{<varint>}\\
  &\text{<varint>} ::= \text{<u8>} | \text{<u16>} | \text{<u24>} | \ldots | \text{<u64>} \\
  &\text{<data>} ::= \text{<[u8]>}
\end{align*}

Where:
\begin{inparaenum}
\item <header> is an optional internal protocol-specific control information.
\item <data> are upper-layer data fragments.
\item  <varint> is a variable-length positive integer whose length-encoding is protocol-specific.
\item The composition may follow protocol-specific ordering constraints.
\end{inparaenum}


A few examples. The Tor end-to-end Relay Cell format~\cite{relay-cell} from a relay perspective:

\begin{itemize}
  \item $H_1 := \text{TCP Header}$.
  \item $H_2 := \text{Relay command}$, $Enc_i = identity$ (no encryption).
  \item $Enc_j = \text{AES-CTR + SHA1}$, and $P_C := \text{<header>} \parallel \text{<data>} \parallel
    \text{<header>} = \left(Recognized \parallel StreamID \parallel Digest \parallel
      Length\right) \parallel \text{<data>} \parallel Padding$. The ordering is fixed.
\end{itemize}

QUIC v1 short header packets may be characterized as follows:

\begin{itemize}
  \item $H_1 := \text{UDP Header}$.
  \item $H_2 := Flags\parallel DestCID\parallel PktNum$, $Enc_i := \text{Selective XOR}$ encrypting
      part of $H_2$ as in Figure~\ref{fig:quicpacket}.
    \item $Enc_j := AEAD$, and $P_C := \text{<chunk>}\parallel \ldots \parallel
      \text{<chunk>}$. Where a chunk is called a frame in QUIC's design defined in
      RFC9000~\cite{rfc9000}. The frame ordering within a QUIC packet is
      arbitrary. The ordering of <field> elements within a frame is fixed.
\end{itemize}

We make the observation that, as currently designed, encrypted transport
protocols cannot be implemented with a contiguous zero-copy interface. 
Indeed, in encrypted transport implementations, the receiving pipeline copies
data to assemble a contiguous buffer of decrypted bytes.
Copies are necessary due
to data fragments being mixed and surrounded by protocol-level control
information in $P_C$. When $P_C$ is decrypted (typically in-place), the
implementation processes some control information, tracks data fragments, and finally delivers
them to the application. If the reading interface to the application is
implemented as providing a contiguous set of bytes, processing data would
require a copy for each data chunk to provide a contiguous stream of bytes.
These copies
may consume non-negligible CPU work, depending on the data location in the
system's memory layout (L2, L3 cache, or main memory).

If the interface to the application offers fragments,
non-contiguous bytes may be passed to the application as they come in, unless
the protocol has a data ordering constraint and fragments are not received in
order, implying copies to buffer them until the ordered fragments come in. In
either case, the application may need to buffer and reassemble the fragments,
involving copies nonetheless.

Changing the structure of the protocol information can safely and securely
unlock contiguous zero-copy receivers in implementations of encrypted transport
mixing protocol control and upper-layer data. Our approach may be applied to
any of these protocols but we expect it to better benefit the ones with a stream
abstraction whose data fragments are eventually meant to be reassembled before
data processing (e.g., QUIC, Tor, TLS1.3/TCP, TCPLS, SSH).

\subsection{Applying Reverso}
\label{subsec:applying_reverso}

\begin{figure*}[!t]
\begin{tikzpicture}[node distance=-\pgflinewidth]

\node[short,fill=black] (a) {};
\labelnode{a}{b}{2cm}{IP Header};
\labelnode{b}{c}{3cm}{Payload};
\node[long,draw=none,fill=none,right=of c,text height=0pt,text depth=0pt,text width=1cm] (d) {$\ldots$};
\labelnode{d}{e}{2cm}{IP Header};
\labelnode{e}{f}{3cm}{Payload};
\node[short,fill=black,right=of f] (g) {};

\draw (a.south east) -- +(0,-0.8cm);
\draw (g.south west) -- +(0,-0.8cm);
\draw[,dotted] (c.north east) -- (e.north west);
\draw[dotted] (c.south east) -- (e.south west);
\draw[<->] ( $ (a.south east) +(0,-0.4cm) $ ) -- node[fill=white] {Kernel Buffer} ( $ (g.south west) +(0,-0.4cm) $ );

\node[above=0.1cm of b](netreadpos) {transport::read()};
\draw[decorate,decoration={brace,raise=2pt}] (c.north west) -- node[above=4pt]
{Copy on read} (c.north east);
\draw[decorate,decoration={brace,raise=2pt}] (f.north west) -- node[above=4pt]
{Copy on read} (f.north east);

\node[short,fill=black, above=1.5cm of a] (aa) {};
\labelnode{aa}{bb}{1.5cm}{Tr. Header};
\chunknode{bb}{dd}{5pt}{2cm}{Data\\ Fragment};
\chunknode{dd}{ee}{5pt}{1.1cm}{Data\\ Footer};
\chunknode{ee}{ff}{5pt}{1.1cm}{Ctrl\\ Header};
\encnode{bb}{cc}{4.85cm};
\labelnode{cc}{gg}{1.5cm}{Tr. Header};
\chunknode{gg}{hh}{5pt}{2cm}{Data\\ Fragment};
\chunknode{hh}{ii}{5pt}{1.1cm}{Data\\ Footer};
\encnode{gg}{jj}{3.55cm};
\node[short,fill=black,right=of jj] (zz) {};
\node[above=0.1cm of bb](trreadpos) {secure\_tr::read()};
\draw[->, thick] (netreadpos) -- ([yshift=-0.1cm]bb.south west);

\draw[decorate,decoration={brace,raise=2pt}] (dd.north west) -- node[above=4pt]
{Decrypted at Position 0} (ff.north east);
\draw[decorate,decoration={brace,raise=2pt}] (hh.north west) -- node[above=4pt]
{Decrypted at Position 100} (ii.north east);

\node[short,fill=black, above=1.5cm of aa] (aaa) {};
\chunknode{aaa}{bbb}{5pt}{2.0cm}{Data\\ Fragment};
\chunknode{bbb}{ccc}{5pt}{1.1cm}{Data\\ Footer};
\chunknode{ccc}{ddd}{5pt}{1.1cm}{Ctrl\\ Header};
\node[long,draw=none,fill=none,right=of ddd,text height=0pt,text depth=0pt,text width=1cm] (eee) {$\ldots$};
\labelnode{eee}{fff}{6cm}{};

\draw[,dotted] (ddd.north east) -- (eee.north west);
\draw[dotted] (ddd.south east) -- (eee.south west);

\node[long, text width=2.0cm, xshift=0.45cm, below=-0.2cm of ccc] (hhh) {Data\\ Fragment};
\chunknode{hhh}{iii}{5pt}{1.1cm}{Data\\ Footer};
\draw[->, thick] (trreadpos) -- ([yshift=-0.1cm]bbb.south west);
\draw[->, thick] ([yshift=0.5cm] dd.north) -- (bbb.south west);
\draw[->, thick] ([yshift=0.5cm] hh.north) -- (hhh.south west);
\draw[decorate,decoration={brace,raise=2pt}] (bbb.north west) -- node[above=4pt]
{len=100} (bbb.north east);


\node[xshift=1cm, right=of g](netbl) {};
\node[xshift=1.25cm, right=of fff](apphl) {};

\node[xshift=2cm, right=of netbl] (netbr) {};
\node[xshift=2cm, right=of apphl] (apphr) {};

\node[yshift=0.8cm, above= of netbl] (nethl) {};
\node[yshift=0.8cm, above= of netbr] (nethr) {};

\node[yshift=-0.8cm, below=of apphl](appbl) {};
\node[yshift=-0.8cm, below=of apphr](appbr) {};

\draw[-, thick] ([yshift=1cm]apphl.north) -- ([yshift=-1cm]netbl.south);
\draw[-, thick] ([yshift=1cm]apphr.north) -- ([yshift=-1cm]netbr.south);

\draw[-, thick] ([xshift=-0.3cm]nethl.west) -- ([xshift=0.3cm]nethr.east);
\draw[-, thick] ([xshift=-0.3cm]appbl.west) -- ([xshift=0.3cm]appbr.east);

\node[xshift=1.5cm, right=of g] () {Transport};
\node[xshift=0.45cm, right=of zz] () {Sec. Transport};
\node[xshift=1.6cm, right=of fff] () {Application};
\end{tikzpicture}
\caption{Data moving up the stack while applying Reverso. A copy of the whole data is
  prevented in between the secure transport and the upper layer. Made
  possible by reversing control information ordering, and by reading the
  decrypted information from right to left instead of the usual reading from
  left to right thanks to the reversed control. Red borders refer to the AEAD encryption of $P_C$.}
\label{fig:reverso}
\end{figure*}
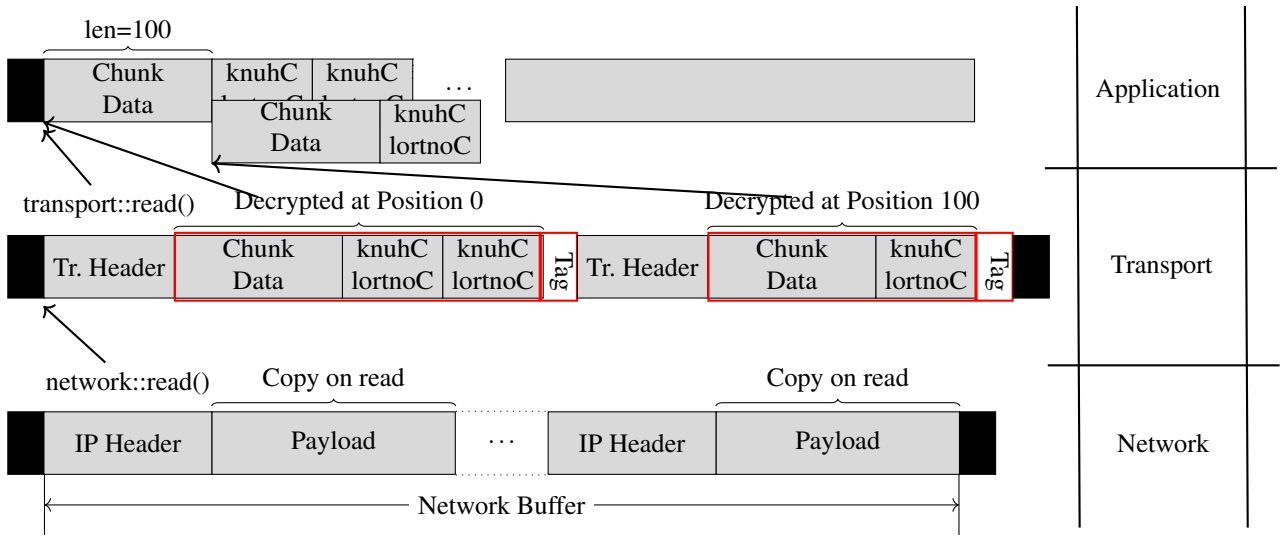


We derive two key principles for protocol specification supporting optimize-out
copies in implementations of encrypted transports but impacting how we process
decrypted bytes, now reading backward once a packet is decrypted instead of
reading forward. We first present the principles, then demonstrate how they can
eliminate data copies.

\mybox[green!20]{ \textbf{Principle 1.} To support backward processing of
  decrypted content in $P_C$, the order of <field> elements in each <header> is
  reversed.  For instance, a <header> with: \texttt{type <u8>}, \texttt{fieldA
    <u16>}, \texttt{fieldB <u64>} becomes: \texttt{fieldB <u64>}, \texttt{fieldA <u16>},
  \texttt{type <u8>}.

}

\mybox[green!20]{ \textbf{Principle 2.} The order and number of <data>
  fragments within a single encryption matter. The data fragment itself must
  always appear as the first element, followed by its associated <header>
  (referred to here as the data footer). Next, any number of control headers in
  any order may be included, limited only by the packet boundary. When multiple
  <data> fragments are written within $P_C$, zero-copy processing at the
  receiver is achievable only for the first fragment. Finally, the ordering of
  <field> elements within all headers, including the data footer, conforms to
  Principle 1.  }

If these two principles are part of the protocol specification, then
implementations may optimize the transport read interface to the above layer
and remove the copy requirement. We may achieve it by exploiting the hidden
copy within the cryptography primitive without changing how cryptography
libraries expose their service (i.e., during decryption, it stores the result
at a destination specified by the caller) and mirroring the read operation
while processing the
decrypted data.

Figure~\ref{fig:reverso} details this process, which generalizes to any
transport protocol involving modern symmetric encryption. In the protocol
implementation, \emph{instead of decrypting in place, we decrypt into the above
  layer's buffer.} There are many options to manage such a buffer. For
the simplicity of our explanations, it can be seen as a contiguous memory space
allocated by the above layer, and for which the encrypted transport layer holds
a reference. 

After decryption, \emph{the implementation can
  safely jump to the end of the buffer} (since the encryption size is always
known), \emph{and process the information backward} until it reaches the
decrypted upper-layer data at the leftmost position. This right-to-left backward processing
is enabled by reversed headers. 

\emph{To construct a contiguous data buffer without copying, the first byte
  position of a packet's data footer can serve as the decryption destination
  address for the next packet.} This approach overwrites the previous packet's
control information with the next data fragment via the cryptographic
primitive's hidden copy, yielding a contiguous data array that can be
passed to the upper layer in a buffer-less manner (without more copying and
buffering for reassembly).

A few critical remarks regarding this technique:

\begin{inparaenum}[1.]

\item If the position of control elements within the encrypted layout is predictable by
  the receiver, i.e., the element's position in $P_C$ is not arbitrary, then
  we may only apply the second principle. It would result in processing the
  control from left to right with the classical wire format, although with a
  data control expressed as a footer rather than a header.

\item If the secure transport protocol offers a multi-stream abstraction, then
  ID information about the leftmost <data> within $P_C$ must be written in $H_2$,
  ideally encrypted and authenticated. This information is decrypted first
  independently of $P_C$ (independent algorithm and independent keys) and
  should then serve as a buffer selection mechanism for $P_C$'s decryption.

  \item To guarantee the efficiency of Reverso, the decryption  of data fragments must happen in
    order. If the packets in the buffer obtained from ``transport::read()''
    are decrypted unordered, then the implementation must save a copy of the
    decryption. To understand why, we may look again at Figure~\ref{fig:reverso}, and assume that the data fragment
    to be decrypted at Position 0 appears second in the transport buffer (not in order). The
    unordered fragment decrypted first at Position 100 would need to be copied
    because part of the control information from the first encryption may
    override part of the data during the decryption process. Mixing unordered
    and ordered packets would result in a partial improvement in proportion to the
    number of ordered sequences. Unordered packets would make the efficiency
    situation for handling them similar to today's protocol implementations.
    However, this can be avoided.  The transport specifications could allow
    the receiver to know the decryption order, for example, by an explicit
    per-stream numbering in the protected transport Header (encrypted) to
    support reordering before decryption, or by using a reliable bytestream
    underneath (e.g., TCPLS).

  \item In the case of a multi-stream encrypted transport, fragments of large streams should
    ideally not be multiplexed within the same packet to guarantee zero-copy for each stream.

  \item A protocol supporting Reverso would have to bump its version
    number, as the structure of the protocol messages changes.

\end{inparaenum}

\subsection{Why Reverso?}
\label{sec:whyreverso}

The approach that we suggest requires revisiting the protocol wire format. For
existing protocol implementations, a minimal supporting implementation means
bumping up the version, having separate code to parse the new wire format
and process control information backward, and respecting the order requirement
for sending data.  Other approaches may be possible, but suffer from serious
downsides.

\noindent
\textbf{Revisiting the Cryptography Interface}.
One other approach than changing the protocol layout could be to revisit the
cryptography interface and give more flexibility to manipulate the
encrypted bytes, for example, to decrypt and process control information in
$P_C$, and then decrypt the data segment into a location depending on the
processed control. However, this would be going against the current security
wisdom gained from the scientific literature showing many catastrophic failures
due to side-channels exploitations in implementations of highly used protocols
in the past with flexible APIs, such as OpenSSH~\cite{albrecht2009plaintext}
for the SSH protocol, OpenSSL for the TLS protocol~\cite{brumley2011remote}, or
MAC-then-encrypt configurations of IPSec~\cite{10.1145/1866307.1866363}.

In consequence of such events, Bernstein \textit{et al}.  explicitly argued for
an atomic interface regarding AEAD capabilities to defend potential padding
oracles~\cite{bleichenbacher1998chosen,
  vaudenay2002security,rizzo2010practical} and potential timing side-channels
that existed against major cryptography libraries in the
past~\cite{bernstein2012security, brumley2011remote}. The standardization of
the AEAD interface, RFC5116~\cite{rfc5116}, accounts for these concerns and
recommends a limited set of input/output driving implementations towards
exposing an atomic interface, which is eventually what most cryptographic
libraries do, therefore making the protocol layout change the reasonable
option.

Furthermore, encrypted transport protocols are designed to be extensible,
usually have a version negotiation procedure, and are not expected to suffer
from ossification~\cite{papastergiou2016ossifying} past the handshake due to their
random looking layout on the wire (once encrypted). As a consequence, applying
Reverso to an existing encrypted protocol should not be a hindrance other than
potentially negotiating and maintaining multiple versions.

\noindent
\textbf{Interface with more than one Destination Buffer}.
For protocols for which the position of the data and control information within
the encryption is predictable by the receiver, one may use an AEAD interface
that supports more than one output buffer, and use the inherent copy
happening during decryption to reassemble the data and move away the control.
In such case, the first principle described in
Section~\ref{subsec:applying_reverso} is unnecessary. The second principle still
applies. For example, the OpenSSL library supports calling multiple times
\texttt{EVP\_DecryptUpdate()}, which we may use to split the
decryption of data and control in two different output buffers.

There are no particular benefits of using multiple destination buffers compared
to using a single destination by overriding the previous packet's control, as introduced
in Section~\ref{subsec:applying_reverso}, neither such an interface is available
for all standardized ciphersuites (e.g., AES-CCM is not supported) or available in many
cryptography libraries, creating a conflict between optimizations,
functionalities, and security for protocol implementations wishing to support
more than one cryptography backend and optional standardized ciphersuites.
Furthermore, this approach requires more meta-data overheads to indicate the
position of the encrypted control.  In some protocols, such as QUIC, the
meta-data could be added to the encrypted header, but such a header does not
exist in every encrypted transport protocol.


%% file: 4_reverso_for_quic.tex
\label{sec:quic_reverso}
The QUIC protocol is an interesting candidate to apply Reverso. Indeed, as the
original QUIC paper wrote~\cite{10.1145/3098822.3098842}, QUIC focuses on rapid
feature development, evolution, and ease of debugging but not CPU efficiency.
Therefore, any efficiency-centered design boost of the QUIC Transport protocol
should be welcome, as it helps with QUIC's core weakness. The adaptation we
suggest follows the design principles in Section~\ref{subsec:applying_reverso}, and
we call the resulting protocol QUIC VReverso.  Importantly, VReverso does not
add, remove, or modify any of the QUIC features, nor do they negatively
impact QUIC's extensibility and security such that the protocol's intended
goals remain untouched. We summarize the changes and then discuss some
important details.

\begin{inparaenum}
\item \textbf{Stream Frame first, then control frames.} QUIC V1 does not
  establish any ordering in a QUIC packet among frames. QUIC VReverso prevents fragmentation
by forcing the reversed Stream Frame to be the first encrypted element (if
any). Other control frames can then be added within the packet in any order.

\item \textbf{Reversed frames.} The wire format of all QUIC frames is
  reversed. Reversing all the fields means that programming the processing
  order in VReverso (from right to left) is the same as QUIC V1 (from left to right) with
  a buffer abstraction that rewinds instead of consuming forward.
  Existing code should need very light adaptation to support the new wire
  format, assuming the existence of a buffer abstraction whose cursor can move
  forward or backward.

\item \textbf{A few more bytes in the short header.} 
  QUIC supports multiple streams and due to UDP, packets may be
  unordered. Because of it, we need meta-data to indicate in which Stream's
  buffer the QUIC payload must be decrypted.  It adds 2 to 8 bytes in the worst
  case in the short header expressing variable integers containing the ID of
  the leftmost Stream frame within the packet and its data offset (2 to 8 bytes
  less for the payload). These added bytes are protected using the QUIC short
  header's protection mechanism~\cite{rfc9001}.

\item \textbf{Multiplexing.} 
The packet may contain other Stream frames with other Stream IDs effectively
multiplexing several streams within a single packet. We \textit{recommend}
in VReverso not to do any multiplexing for multiple large streams to the
peer, as contiguous zero-copy would only be guaranteed for the Stream ID
indicated in the QUIC VReverso short header. Note that QUIC V1 implementations
usually put as much data from a single stream within a QUIC packet, and build
fairness by rotating which next stream fills the next packet, resulting in no
multiplexing in expectation~\cite{marx2020same}. VReverso only gives more
incentive to limit Stream frame multiplexing but does not prevent it for
meaningful use cases (e.g., multiplexing multiple HTTP requests each in their own
stream within a single packet).
\end{inparaenum}

\subsection{Reversed QUIC Frames}

We reverse every QUIC frame to support processing the information from right
to left. For example, the Stream Frame defined in QUIC V1 (RFC
9000~\cite{rfc9000}) to carry application data would have the following
structure once reversed:
~\\

\begin{bytefield}[bitwidth=0.57em]{32}
  \begin{rightwordgroup}{Stream\\ Payload}
    \wordbox[tlr]{2}{App Data} \\
    \bitbox[blr]{32}{}
  \end{rightwordgroup}\\
  \begin{rightwordgroup}{Stream\\ Ctrl}
  \bitbox[blr]{16}{[Length (0 - 16)]}
  \bitbox[blr]{16}{[offset (0 - 64)]}\\
  \bitbox[blr]{24}{Stream ID (8 - 64)}
  \bitbox[blr]{8}{Type (8)}
\end{rightwordgroup}
\end{bytefield}
~\\

Instead of starting from the Type (8) value, and closing with the App Data, the
structure ordering is reversed. A particular detail of attention lies in the variable-length integer definition,
which we also require to reverse within Frames. QUIC commonly uses a
variable-length encoding for positive integer values within most QUIC Frames to
reduce control overheads. The length of the integer is encoded within the
field's two most significant bits in QUIC V1. With Reverso, we require
the length (in bytes) to be encoded within the two least significant bits to accommodate
backward processing of the data. 

Other QUIC frames are then processed with the same logic, consuming all control
information until the reading cursor rewinds to the application data
position, if any. This principle would also apply to all standardized and
ongoing discussed QUIC extensions, since it fundamentally requires respecifying
RFC~9000~\cite{rfc9000}'s frames format.

\subsection{Extended QUIC Short Header}
\label{sec:extended_short_header}

QUIC has the particularity to offer a Stream abstraction to the application
layer, meaning that the application can steer independent data through
different streams. This capability implies independent buffers on the receiver
side. Therefore, regarding Reverso, we must know before decrypting the
payload to which Stream ID the encrypted data belongs, if any. Moreover, since
QUIC reads data from an unreliable channel, we must also indicate the data's
offset. We enable this by securely extending the XOR encryption within the QUIC
short header. The QUIC short packet header is extended as follows:

\begin{center}
  \begin{small}
    \begin{lstlisting}[caption={A QUIC VReverso packet with header protection.
        Added values in red, modification in orange compared to QUIC v1},captionpos=b]
1-RTT Packet {
  Header Form (1) = 0,
  Fixed Bit (1) = 1,
  Spin Bit (1),
  Reserved Bits (2),         # Protected
  Key Phase (1),             # Protected
  Packet Number Length (2),  # Protected
  Destination Connection ID (0..160),
  Packet Number (8..32),     # Protected
  <@\textcolor{red}{ Stream ID (8..32)}@>,               # Protected
  <@\textcolor{red}{ Offset (8..32)},@>                  # Protected
  <@\textcolor{orange}{ Protected Payload (0..72)},@>          # Skipped Part
  Protected Payload (128),   # Sampled Part
  Protected Payload (..),    # Remainder
}
\end{lstlisting}
\end{small}
\end{center}

The short header would now contain the encryption of two new variable-length
integers using similar encoding to the packet number specified in RFC9000's appendix: the Stream
ID of the encrypted Stream Frame if any, or 0 if none, and the data offset.
Shall QUIC officially support Reverso, Stream ID 0 would become reserved to
indicate to the receiver that the QUIC packet only contains control information
(hence, real Stream IDs start at value 1). Since we decided to use at most 4 bytes to
encode the Stream ID, for which 2 bits indicate the length of the offset packed
next, it constrains the peers to have an upper limit of $2^{30}$ streams opened
at any time. In most use cases, applications would limit opening streams to a
much lower value over QUIC's transport parameters during the handshake, since
opening many streams consumes memory to store state information.

Recall from the background Section~\ref{sec:backgroundquic} that QUIC has a
two-level encryption, where part of the header is encrypted using a XOR with a per-packet key.
Therefore, extending information within the QUIC 1-RTT Packet's
header requires extending the key size too, since both informations
are XORed together and must be the same length. In RFC~9001, they call this
key a \texttt{mask}. The mask is derived from the underlying Pseudo-Random
Function (PRF) selected for the current connection. It is either \texttt{AES},
or \texttt{CHACHA20}, producing an output length of a minimum of 16 bytes. The
current header protection applies a XOR operation between the mask and with at
most 5 bytes of header plaintext, up to the \texttt{Packet Number} (at most 5
bytes because the \texttt{Packet Number} has a variable length). In
consequence, we have 11 more bytes of the PRF's output trivially available to
XOR with any extension. To take advantage of Reverso in a multi-stream context,
we are required to extend the short header with at most 8 protected bytes.
Therefore, we truncate the existing one-time pad key to 13 bytes instead of 5.

Eventually, since we add two variable-length integers, we also need modifying
the sampling offset of the encrypted payload used as an input to the PRF. In
QUIC v1, 24 bits are skipped to account for the variable-integer Packet Number
length, which is a guarantee for the receiver to always sample into the AEAD
ciphertext. As a side-effect, it constraints the sender to prepare QUIC packets
with a minimum payload length (in bytes) of

$$\text{min\_payload\_len} = 20 - \text{tag\_len} - \text{pn\_len}$$ to ensure that the
receiver has enough bytes to sample~\cite{rfc9001}. Typical QUIC v1 implementations may
safely set a minimum payload length value of 3 bytes given the current usage of 16
bytes tags.

For QUIC VReverso, we need skipping 72 bits to keep the same guarantee as
QUIC v1 with the two new variable integers. As a side-effect, the relation for
the minimum payload length becomes:
$$ \text{min\_payload\_len} = 28 - \text{tag\_len} - \text{pn\_len} -
\text{stream\_id\_len} - \text{offset\_len}$$
From this relation, a QUIC VReverso
implementation may safely set a static minimum payload length value of 9 bytes given
the current usage of 16 bytes tags (i.e., 9 is the max result the relation may
produce).

%% file: 5_implem.tex
\label{sec:implem}
The QUIC protocol specifications have been recently standardized as
RFC~9000~\cite{rfc9000} after years of efforts starting from Roskin@Google's experimental protocol~\cite{roskind2013quic} building a virtuous circle
between specification guidance and implementation feedback. The circle does
not end with the publication of the RFC. QUIC is designed to resist middlebox
interference to protocol extensibility, allowing the
circle to continue after deployment~\cite{detal2013revealing,hesmans2013tcp,papastergiou2016ossifying}. Our effort for VReverso lies in continuing
this circle. We aim at auditing existing QUIC stacks, leveraging insights for
the potential next iteration of QUIC, and providing an implementation to sustain
our feedback.

\subsection{Existing Efficient QUIC Stacks}

We reviewed 6 QUIC implementations (See Table~\ref{tab:implems}) to estimate
the expected benefits of applying VReverso concerning their implementation
architecture choices. Multiple considerations influence the resulting
architecture and performance. Each project is free to decide what consideration
they care about the most, such as seamless integration in applications or
portability concerns over various OS platforms. Among the existing QUIC
implementations listed by the QUIC IETF working group~\cite{quic-list}, we
looked into the ones that explicitly mention targeting better efficiency or claiming
strong performance, and for which we have experience in the
programming language being used (C/C++, Rust). We can categorize the
implementations into two different architectures. We refer to them as
\textbf{layer architecture}, and \textbf{module architecture}, defined as
follows:

\textbf{Layer architecture.} Implementations within this class follow the OSI
conceptual layering model and aim to provide applications a full abstraction to
exchange secure bytes between two endhosts over the network, with the
additional benefits that the QUIC transport design provides, namely
extensibility, migration, and multiplexing with HoL blocking avoidance. As a
result, applications using this architecture do not have to directly touch the
OS's network abstraction. The QUIC library would do it for them. The interface
of these QUIC libraries typically mimics the insecure layer, which they replace,
``looks'' straightforward to use from the application developer's viewpoint,
and hides to the application the added complexity brought by new multipath
features such as migration. However, portability to different systems and lower
layer IO capabilities is in charge of the QUIC library.

\textbf{Module architecture.} Implementations within this class aim to provide
applications a module to prepare or process QUIC packets, acting as an IO
interface with no expected dependency on the system. As a result, the
application is still responsible for sending or receiving bytes on the network,
which technically offloads portability concerns to the application. QUIC
libraries with this architecture typically have a smaller code footprint,
but a more complex interface and/or lower efficiency than the layer
architecture due to the added inputs/outputs for sending and receiving
information.

\begin{table}[htbp]
  \centering
  \resizebox{\columnwidth}{!}{%
    \begin{tabular}{llcccc} 
        \toprule
        \textbf{Name} & \textbf{Git Head on} & \textbf{Owner} & \textbf{$\approx$ratio memops/dec} & \textbf{Lang.} & \textbf{Arch.}\\
        \midrule
        quiche~\cite{quiche} & bba00c2 &Cloudflare & 36\% & Rust & module \\
        quicly~\cite{quicly} & 8046973 & Fastly & 7\% & C & module \\
        quinn~\cite{quinn} & 36d2b85 & D.Ochtman \textit{et al} & 9\% & Rust & layer \\
        MsQuic~\cite{msquic} & b215b46 & Microsoft & 6.6\% & C/C++ & layer \\
        picoquic~\cite{picoquic} & be0d99e & C. Huitema & 5\% & C & layer \\
        XQUIC~\cite{xquic} & 907be81 & Alibaba & 42\% & C & module \\
        \hline
        \reverso & 83e3c0d & This paper & 0\% & Rust & module \\
        \bottomrule

    \end{tabular}
  }
  \caption{List of reviewed QUIC V1 implementations with performance claims
    that would benefit additional improvements with VReverso. Ratio estimation
    performed through profiling combined with code inspection.}
    \label{tab:implems}
\end{table}



Our Table~\ref{tab:implems}'s review yielded the following observations and insights:

\begin{inparaenum}
\item \textbf{Protocol induced full data copy.} All QUIC implementations require at least one
    copy of all Stream data on the receive path for stream data reassembly, as
    expected from the QUIC protocol design version 1 which is forcing this
    behavior upon implementers due to data fragmentation, as discussed in
    Section~\ref{sec:reverso}.

\item \textbf{Confusing usage of ``zero-copy''.} Some implementations work
    around the memory copy caused by the QUIC design by optionally
    offering an interface with pointers to chunks having the size of the decrypted stream frame
    instead of contiguous data, and call it zero-copy. However, it is not. This
    approach requires the applications to copy into a contiguous array to
    process any information across QUIC data frames, which is essentially
    offloading QUIC's design limitation to the application.

  \item \textbf{Efficiency impact of architectural choice.} Layer architectures are
more likely to yield a more efficient implementation due to controlling
interactions with the OS Network abstraction, controlling buffers and
henceforth avoiding some usage of memory operations. Module architectures on
QUIC V1 may have less control by design if they do not own the data received
from socket IO, which implies memory operations for memory safety reasons to
handle all data fragments and to deal with QUIC's complexity, such as
reordering.  Layer architectures require however much effort to compare on
portability with module architectures, since these implementations have to
write code to handle the different networking abstractions they want to
support. This fact explains why those implementations may typically have a
larger code footprint, assuming everything else is equal (e.g., the amount of
testing coverage).

\item \textbf{Impact of the API choice.} API choices for a module Architecture
may further require an additional copy over the full data path. The memory
ownership of the data pulled from the OS Network abstraction usually belongs to
the application in such architecture implementations. With simple interfaces
between the application and the QUIC library, the QUIC library has to take
ownership by copy of the data. One exception in the \texttt{quicly}
implementation, where a more complex choice for the API and ownership model
supports the library to reassemble the Stream data of different packets and
take ownership with a single memory copy, while other implementations involve
at least two copies and transient allocations/deallocations to achieve the same
goal, but benefit from a more comprehensive interface for the application
developers.
Another exception is
\texttt{quinn} which internally uses \texttt{quinn-proto}, a module
architecture of the QUIC protocol. \texttt{quinn-proto} takes ownership of
incoming data to avoid a second memory copy, at the price to constraint the
embedding of \texttt{quinn-proto} (e.g., \texttt{quinn}) to manipulate received
bytes in heap-allocated segments (stack allocation for data would not be memory
safe).

Designing a networking API is a challenging task with important considerations
for conflicting criteria such as usability, safety, performance and
portability. We discuss observations with more details in Appendix~\ref{appendix:api}, and
give some rationale on how to design an API to offer contiguous zero-copy for a
module implementation using QUIC VReverso.

\item \textbf{Benefit Estimation of VReverso.} We profile each QUIC stack in a
  similar single-stream download scenario in which we continuously reassemble 1MiB of contiguous stream data before dropping it, and we measure the cost of internal
  memory operations caused by the QUIC V1 design relative to the cost of
  decryption within each stack (optimized with AES-NI and PCLMULQDQ hardware
  instructions for AES-GCM in each stack) to provide comparable results between
  stacks based on a common baseline cost. Our measure captures the expected benefits
  VReverso could have on each stack's packet processing: it can be interpreted
  as an approximation for lowering the decryption cost by the indicated
  percentage (Table~\ref{tab:implems}).  Module architectures contain more
  memory operations and management due to data fragmentation and lack of
  control over the Application's buffered data. \texttt{quicly} is an exception
  among module architectures that cleverly solves the second problem at the
  cost of a more complex API and memory model, which brings it to similar
  efficiency to layer architectures while maintaining portability. Its
  conceptor designed \texttt{quicly} to be used primarily with its own HTTP/3
  server implementation, and is not designed for ease of integration.

  A question we consider is whether we can maintain a simple and elegant API,
  such as the one offered by \texttt{quiche} while improving on its efficiency
  limitations. The result of this endeavor would be to unify benefits of both
  architectures: a QUIC VReverso implementation that is portable, efficient
  with no transient allocations/deallocations, no copies due to fragmentation,
  and would be easy to integrate into an Application.
\end{inparaenum}

\subsection{Engineering overview of \reverso}

We present \reverso, our QUIC implementation supporting QUIC V1 and QUIC
VReverso. We base our implementation on Cloudflare's Rust implementation of
QUIC V1 \texttt{quiche}, and follows a module architecture as well. Indeed, \reverso~
has no dependency on the system and has a straightforward interface to
applications, applying our findings discussed in Sections~\ref{sec:reverso},
\ref{sec:quic_reverso}, \ref{sec:implem}, including the contiguous zero-copy
receive API discussed hereafter and further detailed in
Appendix~\ref{appendix:api}. 
As a consequence of this new QUIC protocol version, we
can avoid internal memory management for each packet and two copies of the full application
data inherent to \texttt{quiche} while preserving its elegant API.  To show the
benefits of VReverso, we also modify \texttt{quiche}'s internal HTTP/3 module
to make use of the new API, resulting in the possibility to write an HTTP/3
client and server with increased efficiency, which can process both the HTTP
state machine and HTTP data (i.e., HTTP queries and responses) in contiguous
zero-copy.  Indeed, HTTP/3's stream semantic is a direct mapping to QUIC's
stream semantic, and therefore a design optimization in the QUIC protocol may
directly benefit HTTP/3.

The implementation effort represents about a few hundred lines of change in
\texttt{quiche} to make the protocol compatible with VReverso. Once compatible with
VReverso, the implementation may stay as-is with similar efficiency than
previously. Then, the amount of effort required to make it copy-free is likely
variable from one implementation to another. For \texttt{quiche}, we added
approximately 7k lines of code to adapt its internal, support both protocol
versions, and eliminate copies when VReverso is negotiated. These 7k LoCs
comprise the adaptation of more than $400$ unit tests within \texttt{quiche}
covering all features of QUIC V1 and VReverso to hopefully assert quality. The
resulting implementation and documentation are open-sourced at
\url{https://github.com/frochet/quiceh}.

\subsubsection{\texttt{quiche} and \reverso's interface and ownership model}

\begin{sloppypar}
\texttt{quiche} and \reverso~are module architectures providing an IO interface to the application
for building and processing QUIC packets. We preserved \texttt{quiche}'s
initial choice to provide a straightforward interface with no
system dependency. \texttt{quiche}'s receiving pipeline is an IO interface for
processing contiguous encrypted bytes. The application has to give the
encrypted bytes to the library, and then can expect to read from the decrypted
stream data if a \texttt{conn.readable()} interface returns any `stream\_id'
for a Stream with outstanding data to read. It looks as follows (for
readability, \texttt{quiche} and \reverso's APIs are prepended with their stack
name):
\end{sloppypar}

\begin{small}
\begin{lstlisting}[linewidth=\columnwidth, language=Rust]
// Gives encrypted bytes contained in `buf`
// to QUIC. Returns the number of bytes
// processed or an error code.
fn quiche_recv(
      &mut self, buf: &mut [u8],
      info: RecvInfo,
) -> Result<usize>;
\end{lstlisting}
\hspace{\columnwidth/2 - 0.1cm}\scalebox{1.1}{$\downarrow$} forall readable streams
\begin{lstlisting}[linewidth=\columnwidth,breaklines=true, language=Rust]
// Receives decrypted stream bytes in out
// and returns how much has been written,
// and whether the stream is fin. The
// caller is responsible for providing a
// buffer large enough.
fn quiche_stream_recv(
      &mut self, stream_id: u64,
      out: &mut [u8],
) -> Result<(usize, bool)>;
\end{lstlisting}
\end{small}

\hspace{\columnwidth/2 - 1cm}\scalebox{1.7}{$\circlearrowleft$} Back to quiche\_recv()

\begin{sloppypar}
\reverso~merely modifies and extends the current API by providing a pool of buffers for the
application to use and pass to the \texttt{quiceh\_recv()} and \texttt{quiche\_stream\_recv()}
calls renamed \texttt{quiceh\_stream\_peek()} to capture the fact that this API is not
moving memory out of internal buffers. Furthermore, \reverso~ exposes one more
API call named \texttt{quiceh\_stream\_consumed()} optionally used after peeking
bytes from a stream to mark the data as consumed. In summary, to
process QUIC packets, the application has to perform the following steps:
\end{sloppypar}
\begin{inparaenum}[1.] 
\item Instantiate an application buffer object provided by \reverso. This is
  done once for the whole lifetime of the connection.

\begin{small}
\begin{lstlisting}[linewidth=\columnwidth,breaklines=true, language=Rust]
let mut appbuf = quiceh::AppRecvBufMap::default();
\end{lstlisting}
\end{small}

\item Retrieve the UDP packets using any method of their choice (e.g., using a kernel
  syscall or any kernel bypass such as DPDK~\cite{dpdk}). The chosen method may
  also involve zero-copy of data received from the lower layer, which is
  independent of the optimization VReverso provides.
\item Give the UDP packets containing the encrypted QUIC packets to \reverso's
  \texttt{quiceh\_recv()} interface alongside a reference to the application buffer
  created.

\begin{small}
\begin{lstlisting}[linewidth=\columnwidth,breaklines=true, language=Rust]
// Gives to QUIC encrypted bytes contained
// in `buf`. Returns the number of bytes
// processed or an error code.
fn quiceh_recv(
   &mut self, buf: &mut [u8],
   appbuf: &mut AppRecvBufMap,
   info: RecvInfo,
) -> Result<usize>;
\end{lstlisting}
\end{small}

\begin{sloppypar}
\item The application may then get notified of any Stream ID with outstanding
  available data, and read from the stream using a \texttt{quiceh\_stream\_peek()}
  interface. The application receives a reference to the shared contiguous
  per-stream buffer with outstanding decrypted data to process.
\end{sloppypar}

\begin{small}
\begin{lstlisting}[linewidth=\columnwidth,breaklines=true, language=Rust]
// Receives contiguous bytes from the
// internal stream buffer, starting from
// the first non-consumed byte. Those bytes
// are contained inside the exposed type
// AppRecvBufMap and returned as a
// contiguous array of bytes.
fn quiceh_stream_peek<'a>(
    &mut self, stream_id: u64,
    appbuf: &'a mut AppRecvBufMap,
) -> Result<(&'a [u8], bool)>;
\end{lstlisting}
\end{small}

\item Notify \reverso~with the number of bytes consumed using a
  \texttt{quiceh\_stream\_consumed()} interface.

\begin{small}
\begin{lstlisting}[linewidth=\columnwidth,breaklines=true, language=Rust]
// Tell the QUIC library how much we
// have consumed on a Stream.
fn quiceh_stream_consumed(
    &mut self, stream_id: u64,
    consumed: usize,
    appbuf: &mut AppRecvBufMap,
) -> Result<()>;
\end{lstlisting}
\end{small}
\end{inparaenum}

\begin{sloppypar}
In summary, the application owns the application buffer \texttt{appbuf} and
gives a mutable reference to the \reverso~library, which then sends back a
reference to the underlying contiguous range of outstanding bytes to read for a
given stream (when calling \texttt{quiceh\_stream\_peek()}). That is, the application
and the library share a contiguous memory space for data, and that memory is owned by the
application but managed by \reverso~ through a mutable borrow, and a
signaling mechanism built from the API design (e.g., the
\texttt{quiceh\_stream\_consumed()} interface). Without contiguous zero-copy, we
typically have separated buffers with copies applied to move the data. Our
engineering is similar to lower-layer optimizations, such as io\_uring in Linux or
StackMap~\cite{196184}. The novelty is to observe that such ideas
are now possible with encrypted transport protocols assuming they adapt their
wire layout to match with the design principles discussed in
Section~\ref{sec:reverso}. Note that the current API may be extended to emit
arbitrary sized chunks of internal stream buffers in zero-copy in expectation,
which could be useful for independent multithread processing of streams. This
is a limitation to the current version, for which further engineering may lift
and analyze its impact in a future work.
\end{sloppypar}

%% file: 6_perf_eval.tex
\label{sec:bench}

The following experiments aim to evaluate the efficiency benefits of VReverso
on the received code path with microbenchmarks and a real-world experiment. The
precision required for the microbenchmark evaluation makes such experiments
sensitive to the architecture choice, API choice and language used, making
these results valid only for \texttt{quiche} or any other implementation
sharing the same characteristics.

\subsection{Microbenchmarks}

\textbf{Methodology.}
We use Criterion~\cite{criterion}, an established statistics-driven
benchmarking library for Rust code. Our benchmarking code isolates and compares
the whole processing pipeline of QUIC packets for \texttt{quiche} implementing
QUIC V1 and \reverso~implementing QUIC VReverso using what we could call a
perfect pipe between the sender and receiver (no performance impact from the
pipe) resulting from our setup methodology. We measure the CPU-time spent
processing a given quantity of application data in several microbenchmarks,
hopefully precisely capturing VReverso's efficiency improvement in three
experiments.  That is, the microbenchmarks capture and compare the speed for
both implementations to process QUIC packets, in a reliable and reproducible
way, where every other contributing factor to CPU load than QUIC's processing
code is explicitly not captured (i.e., a perfect pipe assumption).




For all experiments, we report Criterion's best estimate of the processing
throughput within the two libraries with a confidence interval of 95\% over
$5000$ runs for each microbenchmark. For reliable comparison, we use the Linux
utility \texttt{cpupower} to set the processor's min and max frequency to its
nominal value before the experiment, and we use the Linux utility
\texttt{taskset} to pin the experiment over a single core. All experiments are
single-threaded. We use a maximum datagram size of $1350$ bytes, to obtain
measurements matching the datagram size expected to work best in the real
world, as recommended by Google~\cite{google-datagramchoice}. Note that while
we compare efficiency based on a throughput metric, those experiments are not
designed to maximize throughput, but to reliably compare two designs (QUIC V1,
and QUIC VReverso on a portable QUIC implementation forked from
\texttt{quiche}).


\begin{figure*} 
  \centering
  \begin{subfigure}[t]{0.335\textwidth}
    \includegraphics[width=\textwidth]{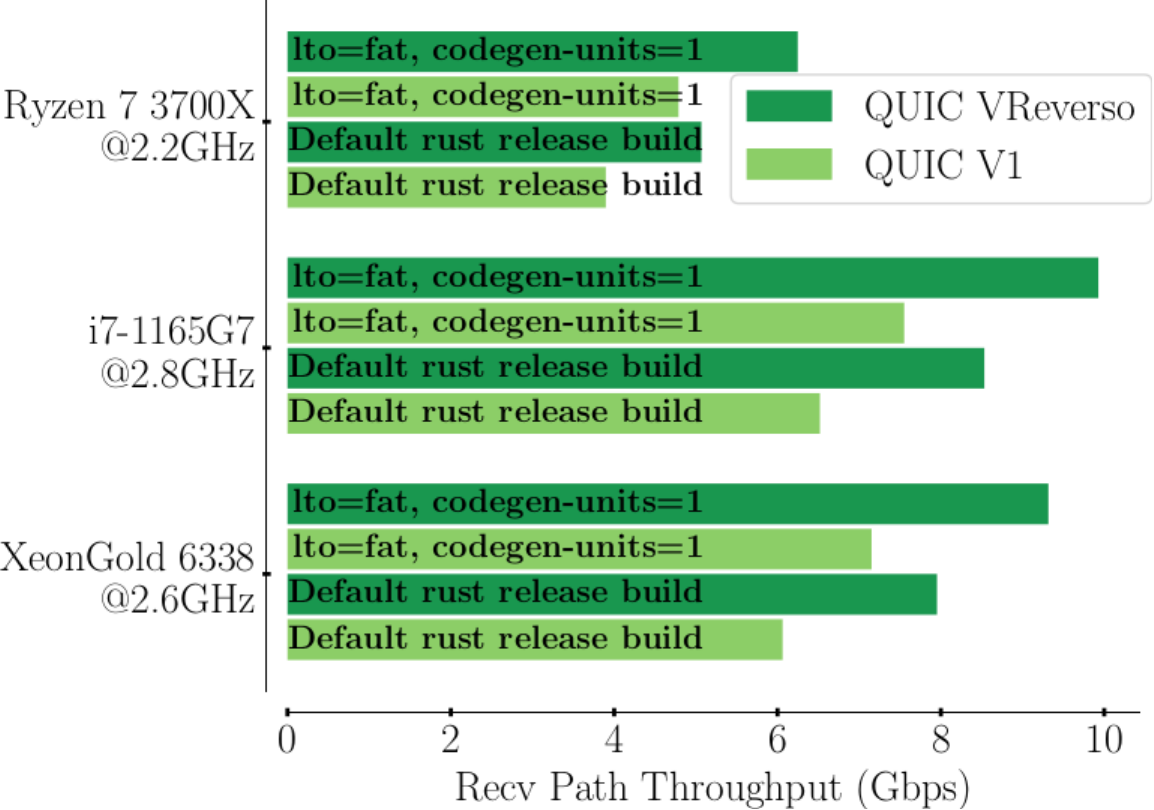} 
  \caption{Packet processing QUIC V1 compared to QUIC VReverso.}
  \label{figure:early_results}
\end{subfigure}
\hfill
  \begin{subfigure}[t]{0.305\textwidth}
    \includegraphics[width=\textwidth]{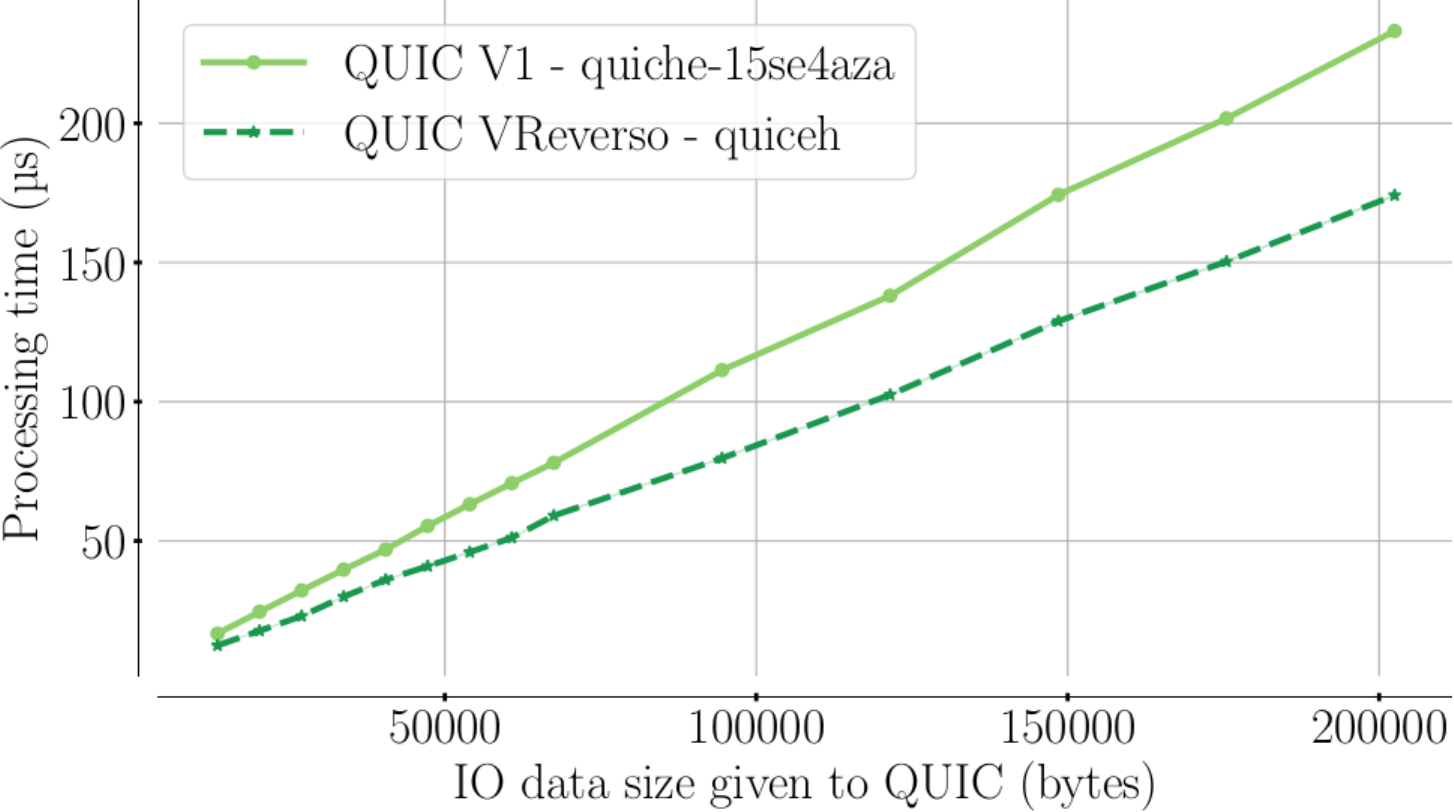}
    \caption{Impact of IO data size.}
  \label{figure:buflen}
\end{subfigure}
\hfill
  \begin{subfigure}[t]{0.335\textwidth}
    \includegraphics[width=\textwidth]{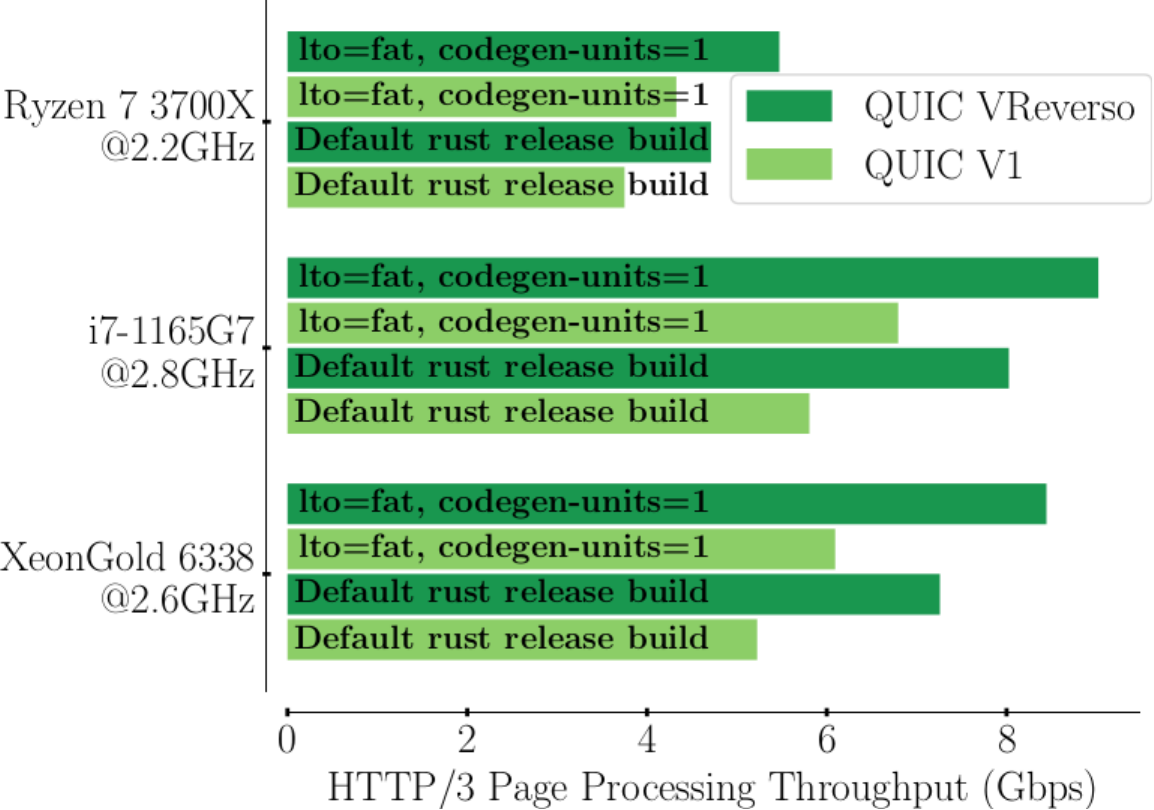}
  \caption{HTTP/3 processing efficiency of a 2,5MB page spread into 80 HTTP/3
    request/response streams.}
  \label{figure:http3}
\end{subfigure}
\caption{Microbenchmarks comparing QUIC to QUIC VReverso on a Module
  architecture implementation design. Results averaging 5000 runs with C.I. of
  95\% (interval too small to appear).} \label{fig:microbenchmarks}
\end{figure*}

\textbf{(Fig.~\ref{figure:early_results}) Benchmarking on several processors architectures.}
Fig.~\ref{figure:early_results} displays results evaluating the processing
speed of a full initial congestion window (10 packets). Applying VReverso, we
were able to optimize out two memory copies of all the application Data.

Compared to the baseline,
VReverso is increasing the throughput of QUIC packet processing by $\approx 30\%$
for all three tested processors using Rust's default build configuration with ThinLTO~\cite{47584}. The
improvement is consistent with$\approx 30$ to $31\%$ higher efficiency when we
compile both libraries using fat LTO. This is an interesting result showing that
VReverso's benefit is directly cumulative to efficiency benefits from compiling
with a better optimization level. Altogether they improve packet processing by
$\approx 52\%$ on \texttt{quiche}'s default build, across all tested
configurations on their base rate frequency.

Interestingly, our results show that the efficiency gain from VReverso is stronger
than cross-module Link Time optimization over all \texttt{quiche}'s crates dependencies
(fat LTO). This compiler optimization can lead to significant benefits, as
depicted here, but it is rarely used by large projects with many dependencies
due to a significant increase in build times. 


\textbf{(Fig.~\ref{figure:buflen}) Varying the number of packets fed into the QUIC library.}
Typical usage of either \texttt{quiche} or \reverso, as per their design
architecture, implies the application reading data from the OS Network
abstraction, and then feeding it to the library for processing. The quantity of
data read at once is a value up to the maximum OS's receive buffer size,
potentially containing multiple datagrams and multiple QUIC packets. In our
experiments, the maximum receive buffer size is Linux's default value of
$212992$ bytes, meaning that the application is expected to give a quantity of
data in the interval $[0, 212992]$ to \texttt{quiche} or \reverso's
\texttt{recv()} interface. Indeed, the expectation is that after draining the
data from the network, the application would then directly pass it to the QUIC
library (there is no incentive to hold on to it and pass more than 212992 bytes,
unless the application tweaks the socket option SO\_RECVBUF).
Figure~\ref{figure:buflen} captures this assumption by benchmarking multiple
lengths of buffered data passed to the QUIC library, those lengths being
themselves a multiple of the UDP maximum datagram size we're using ($1350$
bytes). The goal of this experiment is to show the expected efficiency gain
across this range of potential events, giving us a proxy of the expected
efficiency of \reverso's packet processing in the real world if applications start using our
design. Our measures show a decrease of data processing time of $\approx 26\%$
across the x-axis. Note that increasing the size of the kernel's recvbuf is
then expected to increase throughput on large bandwidth, since there is no
negative effect on the QUIC receiver.

\textbf{(Fig.~\ref{figure:http3}) Powering up HTTP/3 with VReverso.}
Our implementation also exposes a zero-copy API for processing HTTP/3 requests
or responses. This implementation is an extension of \texttt{quiche}'s internal
HTTP/3 module, and modifies it to directly pass to the application
\reverso's contiguous data. Internally to the HTTP/3 module, all control
information (e.g., handshake-related information) is also processed in
zero-copy directly from \reverso's contiguous stream data. Since HTTP/3 is
technically an application protocol directly built atop QUIC, then any
efficiency improvement in the QUIC protocol would translate to an opportunity
for an efficiency improvement in the HTTP/3 implementation. In our
implementation, we support exposing HTTP/3 Data frame of arbitrary length in zero-copy.
The HTTP/3 protocol itself is untouched but requires negotiating QUIC VReverso
to benefit from the contiguous zero-copy.  Demonstration of this claim is given in
Figure~\ref{figure:http3}.  Reverso's optimization directly translates to
HTTP/3, and we report $\approx 38\%$ data throughput improvement on both the
XeonGold and the i7 processors.  We, however, noted that the efficiency
improvement was $\approx 26\%$ on the Ryzen, and would increase to $\approx
30\%$ by batching a smaller quantity of UDP packets at once to \reverso. This
result shows that the Ryzen, in contrast to the intels, is more sensitive to
different input sizes in the library. A smaller L2 cache for the Ryzen may
explain the result.

Note that the interface of the H3 module to the application also have minor changes to
support the contiguous zero-copy abstraction. We adapted the original \texttt{quiche}'s
HTTP/3 client and server demonstration application to support \reverso. A
demonstration server running both the new QUIC protocol and QUIC V1 underneath
HTTP/3 is available at \url{https://reverso.info.unamur.be}

\subsection{Real-world benchmarks}

\begin{table}[h]
    \centering
    \setlength{\tabcolsep}{5pt}
    \renewcommand{\arraystretch}{1.2}
    \newcolumntype{Y}{>{\centering\arraybackslash}X|}
    \begin{tabularx}{\columnwidth}{@{}l|YYYYYYY@{}}
      $\downarrow$ C, $\rightarrow$ S & BE & FR & DE & CA & SG & AU & PL \\ \hline
        BE & \cellcolor{gray!25}  & 99.74 & 99.76 & 99.52 & 99.72 & 99.52 & 99.73 \\ \hline
        FR & 90.11 & \cellcolor{gray!25}  & 99.96 & 99.94 & 99.90 & 99.90 & 99.92 \\ \hline
        DE & 89.36 & 99.96 & \cellcolor{gray!25}  & 99.88 & 99.89 & 99.89 & 99.91 \\ \hline
        CA & 96.90 & 99.90 & 99.90 & \cellcolor{gray!25}  & 99.84 & 99.92 & 99.96 \\ \hline
        SG & 94.30 & 99.90 & 99.88 & 99.92 & \cellcolor{gray!25} & 99.92 & 99.92 \\ \hline
        AU & 98.39 & 99.92 & 99.89 & 99.94 & 99.90 & \cellcolor{gray!25}  & 99.96 \\ \hline
        PL & 96.00 &  99.85 & 99.84 & 99.93 & 99.79 & 99.92 & \cellcolor{gray!25}  \\ \hline
    \end{tabularx}
    \caption{Ratio (in \%) of ordered QUIC packets from HTTP/3 transfer with
      VReverso in between vantage points acting as both server (S) and client (C) from countries in
      America, Europe, Asia and Oceania (100 Mbps Internet connectivity).}
    \label{tab:reordering}
\end{table}

\noindent
\textbf{Reordering in the wild}.
The microbenchmark results assume a perfect pipe where no reordering happens.
The perfect pipe experimental setup was designed for a reliable and precise comparison of the impact of VReverso
over a typical QUIC implementation. Of course, the Internet is not perfect.
Therefore, we aim to evaluate the likelihood of packet reordering on the
Internet using our new QUIC version. As previously explained in
Section~\ref{sec:reverso}, a packet not arriving in order has the consequence of implying a copy.

To the best of our knowledge, a single research paper looked at UDP reordering
in the wild in 2004~\cite{zhou2004reordering} stating UDP reordering as a rare
event. Since it is reasonable to assume that the results could be different
after 20 years, we aim at verifying whether reordering is still rare today. We
set up a few vantage points in North America, Europe, Singapore and Australia
and we performed 20 HTTP/3 VReverso downloads of 2.5MiB content between each
peer, each acting as a client, and then as a server. We record whether the
packets were received in order, and for which contiguous zero-copy as
introduced in Figure~\ref{fig:reverso} was performed thanks to our protocol
changes.

Table~\ref{tab:reordering} shows that an overwhelming fraction of the data
transfer is in order and does happen in zero-copy. An anomaly was, however, recorded from the
Belgian vantage point, where the fraction of ordered packets oscillates between
$\approx 89\%$ to $\approx 98.4\%$ depending on the peer. We tracked down the
problem to a potentially poor DPI performed by a Firewall close
to the BE vantage point. Using a machine in this local network
that does not go through the Firewall does not create the anomaly, and gets us
$>99.5\%$ of ordered packets against each other locations.
Research works improving the efficiency of networking systems by preserving packet
ordering within NICs and middleboxes could be of importance~\cite{285094} for
the future of encrypted protocols.

These results, while not standing for the global Internet, do however suggest
that most QUIC packets should be ordered in the wild, and not necessarily
linked to the distance, due to the congestion control kicking in and reducing
bursts of packets on lossy links. We observed fewer reordering issues through
the Firewall over high-distance QUIC connections, which may be explained by
congestion events we observed  over the high RTTs connections, which may be
reducing the DPI cost on the Firewall and the resulting impact on ordering.

\noindent
\textbf{1 Gbps HTTP/3 download}.
\begin{figure}[t]
  \centering
  \includegraphics[width=\columnwidth]{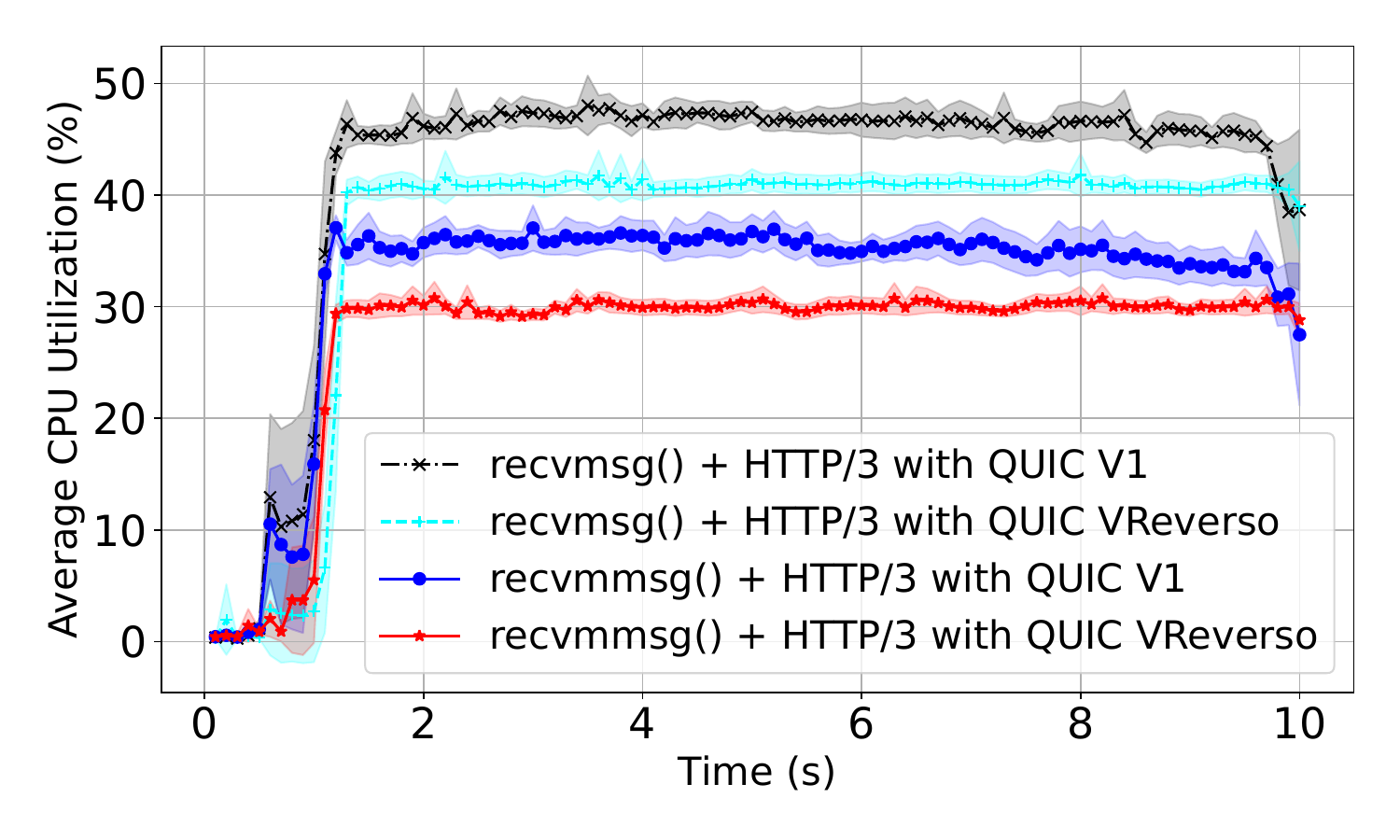}
  \caption{HTTP/3 download CPU utilization with $95\%$ C.I. monitored on a
    receiver using a i7-1165G72.8GHz with frequency scaling disabled and maxing
    out 1 Gbps link on a process using a single core and single thread,
    pinned to a single CPU.}
  \label{fig:http3dl}
\end{figure}
While the microbenchmarks measure the QUIC packet processing improvement thanks
to VReverso, this evaluation gives more insights about expected benefits with a real
network. For each of the experiments, we perform 20 HTTP/3 downloads for 10
seconds, saturating a Gbps link with a client using an i7-1165G7 processor with a
frequency set to its nominal value (2.8GHz), preventing CPU frequency scaling
and therefore improving the reliability of the CPU utilization comparison.
We measure CPU cycles consumed by the QUIC client every 100ms and display the
average over the 20 downloads with a 0.95 confidence interval and compare
Reverso's improvement to the otherwise known and significant improvement of UDP
syscall extensions to process/send batches of UDP packets.

Indeed, a known efficiency problem of QUIC is not QUIC itself, but its carrier: UDP.
Every QUIC packet is transported within an UDP packet, and previous research
papers~\cite{10.5555/2342821.2342830, tcpls-conext, 10.1145/3589334.3645323} as well as industrial
reports~\cite{industrial-quic-report} have pointed to the kernel's interface
(\texttt{recvmsg()} and \texttt{sendmsg()}) sending a single UDP message per
call as being an efficiency bottleneck. Using the kernel's syscall extensions
\texttt{recvmmsg()} is known to have a major impact on
improving QUIC's perceived efficiency (and \texttt{sendmmsg()} for specific cases). However, usage of these syscalls is not
necessarily widespread due to a variety of reasons. In some languages, like
\texttt{Rust}, it is not yet part of the standard API, so using them may incur
significant implementation efforts if one wishes to guarantee integration
stability (much more labor than being compatible with VReverso) or may be delayed until
support is eventually added. We add support of \texttt{recvmmsg()} in our
HTTP/3 client using \texttt{quinn-udp}, a third-party open-source socket
wrapper supporting the syscall extensions.  We compare VReverso's benefits to
the benefits of the syscall extensions that serve as a baseline to appreciate improvement. As
Figure~\ref{fig:http3dl} shows, VReverso's benefits are roughly similar to half
of the benefits of receiving/sending multiple datagrams in one syscall. These
experiments also show that those benefits cumulate. That is, the more the
transport aspects of QUIC packets are improved, the more important VReverso's
impact becomes compared to the baseline.  As QUIC's surrounding environment
improves through the years, VReverso's perceived benefits to data receivers are
expected to increase.

\noindent
\textbf{Limitations}.
The evaluation we provide has required writing the necessary tooling for the
new protocol and new QUIC library \reverso~including an HTTP/3 client/server
instrumenting \reverso, with the support of \texttt{recvmmsg()}. These tools, however, offer a limited vision of the
expected efficiency improvement real clients would obtain among the myriads of
potential applications using QUIC VReverso. First, the  efficiency benefits of
VReverso depend largely on one implementation to another. Comparing
several implementations on microbenchmarks or a real-world measure of CPU consumption becomes
meaningful if VReverso is implemented into them as well and compared to their
own baseline to better understand the expected overall improvement.  Second, compared to related works evaluating
QUIC~\cite{10.1145/3589334.3645323} on real-world applications, we cannot yet
do any integrated analysis on websites browsing or video streaming. Doing such
analysis would require integrating \reverso~into an existing browser and web
server over existing services. The current
evaluation comparing \reverso~to \texttt{quiche} as well as the
benefits approximation from Table~\ref{tab:implems}
hopefully hints at the benefits of extending QUIC
and starting integration efforts. In the future, we
may hopefully evaluate VReverso within a myriad of
deployed applications.

\noindent
\textbf{Energy Efficiency}. Reverso is a software optimization, and as in any
optimization, it may lead to energy savings in some context. We expect, however,
these savings to be the exception rather than the norm. In mobile-based
communication, most of the energy is spent on powering antennas during I/Os
communications, and improvements focus on being careful with I/Os
triggers~\cite{qian2011profiling} rather than CPU savings. The same goes with
radio-based IoTs. In wired-based setups, such as desktop and servers,
while it is true that Reverso can bring energy savings while processing packets
of say, an HTTP/3 response, it needs to be put in perspective of what the
server was computing to produce the response. If the initial request triggers
several GPUs to compute an answer, which is eventually sent back to the
client, does our optimization matter considering a holistic view of energy
spending? It is likely negligible, considering a holistic view of energy
consumption.  Reverso is helpful in cases where the transfer bottleneck is the
CPU, which will cause an increase in transfer speed at constant energy
consumption if our optimizations are used. Reverso has no claim to help with
any ecological problem, as optimizations without reductions in service
consumption do not help with sustainability~\cite{segovia2023efficiency}.



%% file: 7_sec.tex
\label{sec:sec}

QUIC VReverso requires careful usage of header information before AEAD decryption,
considering an on-path adversary manipulating the packet, since any malformation would
be discovered later at the AEAD decryption stage, where the header information is fed into
the cipher as associated data.

In QUIC V1, the AEAD algorithm ensures that any modification to either the QUIC
header or payload bits will result in a decryption error. Meanwhile, the
receiver has no usage of information from the decrypted header before the AEAD
decryption succeeds.

In QUIC VReverso, part of the protected header information might be
needed before payload decryption, for instance, when the header contains
a new Stream ID requiring a new stream buffer prior to 
decrypting the QUIC payload into it. This early usage of the metadata contained within
the decrypted header demands careful implementations to prevent two specific
issues:
1) A potential side-channel exposing the stream state information to on-path
adversaries. 2) A memory safety violation where previously valid decrypted
data could be overwritten by a manipulated offset from a next packet. Both scenarios
can be prevented with proper implementation safeguards. We discuss these two
topics hereafter.

\subsection{Side-channels}
\label{subsec:sidechannel}
Encrypted transport protocols design and implementations mixing encryption of
control information and payload have a literature body of attack
papers~\cite{10.1007/3-540-46035-7_35, 10.1007/978-3-540-45146-4_34,
  10.1007/978-3-642-25385-0_20, albrecht2009plaintext, 10.1145/1866307.1866363,
  Paterson2012PlaintextRecoveryAA, 6547131, 10.1007/978-3-662-49890-3_24} involving timing side-channels allowing
partial plaintext recovery by an on-path attacker without knowledge of the
secret key. These attacks abuse broken cryptography constructions usually involving
MAC-then-Encrypt with a CBC encryption mode. Moreover, to succeed in exploiting
the cryptography vulnerability, the attacker needs several stepping stones. The
first one is the existence of a timing side-channel within the processing
packet implementation, impacting the time to decrypt data and controlled by the
on-path attacker. For example, it could be a partially decrypted but not
authenticated control field such as a ``length'' information that is used to
decrypt the remaining of the packet. A second stepping stone is the ability to
observe endhost timing side-channels from the on-path network attacker
position. Usually, it involves distinguishing the endhost's reaction 
to packet processing errors by observing emitted network messages.

Regarding QUIC VReverso, the cryptographic construction rationale from QUIC V1 is
untouched: we have two independent decryption algorithms using different keys
and each is designed to be atomic, i.e., we decrypt the whole header at once with
the decryption header algorithm, and then we decrypt the payload at once
with the payload decryption algorithm. Within the payload, any decrypted control
information does not impact the decryption process. There is no known cryptography
vulnerability to exploit within each of these algorithms.

A timing side-channel may, however, exist if the implementation handles a new
Stream ID differently than an existing Stream ID, e.g., by allocating a new buffer
before decrypting into it, finding an integrity error, and then freeing the new
buffer. \reverso~uses a recycling buffer strategy, where there's always a
pre-allocated buffer available for a potential new stream. If a decryption
error is detected within a fresh stream, the buffer is recycled and ready for
any other new stream. Although, even if a QUIC VReverso implementation does not
follow such a strategy and has a timing side-channel, this is not particularly
useful for the attacker for two reasons 1) it may only allow to guess some
information about the state of the QUIC connection, e.g., the number of opened
streams. 2) exploiting the timing side-channel requires an observable reaction
on error visible on the network. The QUIC protocol is silent on error; nothing
is sent on the network. It is unlikely an attacker could observe the timing
side-channel even if one exists.



\subsection{Memory safety}

Assuming the decoded Stream ID already exists, the receiver must
not attempt to decrypt into the application's stream buffer at an offset lower
than the highest contiguous received offset. Such programming would be
unsafe for VReverso in the case where the offset value is manipulated and if
the AEAD decryption writes at the destination address before the tag is checked
(some AEAD implementations do). Therefore, if the decoded offset is lower than
the highest contiguous offset, we must attempt decryption in place.
Receiving an offset lower than the current highest contiguous offset may happen
in non-adversarial but rare usage of QUIC due to spurious retransmissions of
received data not acknowledged fast enough to the sender.  In case the
integrity succeeds, the packet would be acked but the stream frame dropped (it
is spurious: we do not need the content, but we need to tell our peer that we
received it). In case integrity fails, it is a malformation, the in-place
decryption keeps the destination buffer safe from any incorrect write, and we
may drop the packet.  For any decoded offset greater or equal to the current
highest contiguous offset, AEAD decryption does not raise any safety issues,
since the implementation receiving any out-of-order data (that is, any offset
$>$ highest contiguous offset) has to make a copy of the decryption for another
safety reason (see critical remark (3) of
Section~\ref{subsec:applying_reverso}), and this copy is what must be delivered
to the application eventually.

Our \reverso~implementation implements the described stream memory safe
behavior.

%% file: 7_relatedwork.tex

\noindent
\textbf{Applicability of Reverso.}
Our work directly applies to academic and industrial QUIC and HTTP/3
extensions~\cite{9433515}, such as Datagram~\cite{rfc9221}, Multipath
QUIC~\cite{de2017multipath, quic-multipath, pquic}, FEC~\cite{michel2022flec,
michel-thesis}, Proxying~\cite{kuhlewind2021evaluation} and would contribute to
more efficient VPNs in general. The principles required for taking advantage of
Reverso's optimizations described in Section~\ref{sec:reverso} are not protocol
specific and may be added to any of them, including many proposals from
academics~\cite{nowlan2012fitting, tcpls-hotnets, tcpls-conext,
  jadin2017securing}, and existing privacy-centered solutions such as
Tor~\cite{tor}. The implementation of these protocols, if Reverso is applied,
should also perceive some level of efficiency improvement. 

It is also possible to directly apply our implementation technique on TLS 1.3
userspace implementations providing an IO interface (e.g.,
Rustls~\cite{rustls}, GnuTLS~\cite{gnutls}, WolfSSL~\cite{wolfssl}) without
changing the protocol, since TLS 1.3 is not mixing encrypted control
information before adding data and the position of encrypted control after the
data is predictable by the receiver. As a consequence of the design of TLS 1.3,
the principle 1 (Section~\ref{sec:reverso}) is unneeded, and Principle 2 is
already met by TLS 1.3 specifications, since the protocol has a single control chunk of 1 byte
encrypted next to the data (ContentType).  Existing TLS implementations may then directly
exploit the inherent copy existing in decryption as depicted in
Figure~\ref{fig:reverso} to offer an interface to applications with contiguous
zero-copy. To the best of our knowledge, none of the aforementioned
implementations yet offer contiguous zero-copy, while some of them do offer a
packet-based zero-copy interface (i.e., limited to the size of a TLS record),
indicating a desire to offer zero-copy but maybe missing the insights discussed
in Section~\ref{subsec:applying_reverso} to realize it at full strength.

\noindent \textbf{Zero-copy and stack engineering.}
While this research paper explains how to design an encrypted protocol to
eventually support engineering zero-copy in the implementation, research
redesigning the stack for lower layer transfer has been a hot topic in
system communities~\cite{10.1145/3458336.3465287,
10.1145/3458336.3465283, 10.1145/3600006.3613137}.
Netmap~\cite{10.5555/2342821.2342830} is a FreeBSD
framework bridging NICs to network APIs removing existing syscalls, data
copies, and meta-data management overheads to bring data up the stack more
efficiently. StackMap~\cite{196184} further improves netmap's engineering for
TCP/IP. Such effort may be combined with the contribution of this research, as
\reverso~is designed to be system independent and happens on the layer above
those contributions. A QUIC stack (now unmaintained),
quant~\cite{eggert2020towards} uses such architecture design but applied to
UDP/IP instead of TCP/IP. Further work could cumulate efficiency benefits
by applying lower-layer stack optimizations before handling packets to a QUIC
implementation supporting VReverso.



%% file: 8_ccl.tex
%

This paper introduces the idea that protocols mixing encrypted control and data
should be specified with two design principles coined as ``Reverso'' to take
advantage of the hidden memory copy inherent to cryptography primitives for
message reassembly across packets. Designing future encrypted protocols with
Reverso offers the opportunity for implementers to obtain higher efficiency by
offering a contiguous zero-copy interface to the upper layer, which is
otherwise impossible due to a combination of inflexible cryptography APIs (for
good reasons) and protocol layout fragmentation in today's encrypted protocol
design space.  We discuss these claims by describing the fundamental causes of data
fragmentation requiring copy overheads, then specifying principles that an encrypted
transport protocol should satisfy to prevent fragmentation and by redesigning
the QUIC protocol with those principles. We open-source \reverso, a full
implementation demonstrating the idea to the QUIC community using a portable
software architecture with a reading API providing contiguous bytes. \reverso~
on VReverso bypasses two full data copies in its processing pipeline compared
to v1 which were inherently caused by a combination of the QUIC v1 layout and
\texttt{quiche}'s desire for portable code with a streaming API.  Importantly,
our VReverso implementation maintains \texttt{quiche}'s portability and
contiguous reading API (now in zero-copy). Other QUIC implementations owning
different software engineering choices may see their improvement with Reverso
to be lower or higher depending on the stack. In all cases, an implementation
of this optimization should lead to some efficiency benefits for providing
contiguous bytes to the application. For some stacks or other existing protocols
and implementations, the benefit may not be worth the effort, but this choice
is left for the implementer to make once the fundamental principles are applied
to the protocol's wire image to support it.

Reverso hopefully raises interest to many use cases carrying streams of data
that should benefit from such optimization, such as VPNs, IP tunnels,
high-speed datacenter transfer, video streaming, QUIC-based privacy-enhancing
technologies and many more. The insights discussed within this research are not
limited to QUIC and should raise interest for other encrypted protocol designs
as well, while the exact benefit for them remains unknown.

%% file: 1_appendix.tex
\subsection{Open Science}
All results within this paper are reproducible with \reverso~ open-sourced and
publicly available under the BSD-2-Clause License granting commercial use,
modification, distribution and private use of this research. The code is
available for review and instructions to reproduce the core results
(Figure~\ref{fig:microbenchmarks}, Table~\ref{tab:reordering} and Figure~\ref{fig:http3dl}) are available
at~\url{https://github.com/frochet/quiceh/REPRODUCE.md}.

\subsection{API Design Rational for Zero-Copy}
\label{appendix:api}

Designing a networking API is a challenging task. Usability, safety,
performance, portability, and stability are usual considerations to interface
library features to the application, leading to much variability in
implementations. A natural choice for a module architecture reading bytes from
a stream could be composed of two API calls. This is exemplified in
\texttt{quiche} with the following API functions:

\begin{small}
\begin{lstlisting}[linewidth=\columnwidth,breaklines=true, language=Rust]
// Gives encrypted bytes contained  in `buf`
// to QUIC. Returns the number of bytes
// processed or an error code.
fn recv(
      &mut self, buf: &mut [u8],
      info: RecvInfo
) -> Result<usize>;
\end{lstlisting}
\hspace{\columnwidth/2 - 0.1cm}\scalebox{1.1}{$\downarrow$} forall readable streams
\begin{lstlisting}[linewidth=\columnwidth,breaklines=true, language=Rust]
// Receives decrypted stream bytes in out
// and returns how much have been written,
// and whether the stream is fin. The
// caller is responsible for providing a
// buffer large enough.
fn stream_recv(
      &mut self, stream_id: u64,
      out: &mut [u8],
) -> Result<(usize, bool)>;
\end{lstlisting}
\end{small}

\hspace{\columnwidth/2 - 0.52cm}\scalebox{1.7}{$\circlearrowleft$} Back to recv()

Applications using such a module would then typically read from the network, then
give their buffered bytes to the QUIC module though the
\texttt{recv()} API, and eventually obtain contiguous decrypted
stream bytes from the \texttt{stream\_recv()} API call. This choice provides
clear usability for the application but forces the QUIC library developers to
have two copies of the whole application data. One of these copies is caused by
reassembling the Stream frames of different QUIC packets, as we already
previously discussed, and which we address thanks to VReverso. The other
copy is caused by an equally subtle fact: \textit{The QUIC connection's implementation abstraction may
  outlive the input `buf' buffer and output `out' buffer}. This is a safety problem
caused by the API choice.  Since both `\texttt{buf}' and `\texttt{out}' are
owned by the application, and since QUIC packets can be unordered, the QUIC
module may require several calls of \texttt{recv()} before being
able to reorder the QUIC data and deliver them through
\texttt{stream\_recv()}. Because the QUIC module does not own
`\texttt{buf}' and `\texttt{out}' and may outlive them, safety is only guaranteed by copying
`\texttt{buf}' 's content at each \texttt{recv()} call. The copy may happen before decryption, or after in-place
decryption. The latter option is however better performance-wise, as we would
only copy application data and no other information. This strategy is what
Cloudlare's \texttt{quiche} implementation does.

\texttt{quicly}, another module Architecture implementation of QUIC has a
clever efficiency optimization to solve this safety problem. Their idea seems
to essentially design the API to address both the safety issue and message
reassembly in a single memory copy. To achieve this, \texttt{quicly} does not
take in an \texttt{out} buffer pointer like \texttt{quiche} does, but chooses to
own the stream output buffer and expose it to the application when needed.
The application has to configure a callback function \texttt{on\_receive()}
called on the reception of each QUIC packet. On each of these calls, the
application must further call one other \texttt{quicly} API to process the
received decrypted data. This is where the copy happens; i.e., \texttt{quicly}
takes ownership of the in-place decrypted bytes and reassembles through a
single copy in an internal contiguous stream output buffer. The application can
then optionally call further APIs to: 
\begin{inparaenum} 
\item  Obtain a pointer to \texttt{quicly}'s internal stream output buffer. 
\item Tell \texttt{quicly} how many bytes have been consumed from the internal
  stream output buffer.
\end{inparaenum}
We then have a trace of API calls as follows for the application:

\begin{small}
\begin{lstlisting}[linewidth=\columnwidth,breaklines=true]
// Gives encrypted bytes contained in
// `packet'. Returns 0 or an error code.
int quicly_receive(
    quicly_conn_t *conn,
    struct sockaddr *dest_addr,
    struct sockaddr *src_addr,
    quicly_decoded_packet_t *packet
);
\end{lstlisting}
\hspace{\columnwidth/2 - 0.1cm}\scalebox{1.1}{$\downarrow$} callback on\_receive()

\begin{lstlisting}[linewidth=\columnwidth,breaklines=true]
// Takes ownership and reassembles the
// decrypted stream data located within
// the decoded packet into an internal
// stream output buffer.
int quicly_streambuf_ingress_receive(
    quicly_stream_t *stream, size_t off,
    const void *src, size_t len
);
\end{lstlisting}

\begin{center}\scalebox{1.1}{$\downarrow$}\end{center}

\begin{lstlisting}[linewidth=\columnwidth,breaklines=true]
// Gets the internal quicly stream buffer
// type holding contiguous stream bytes.
ptls_iovec_t quicly_streambuf_ingress_get(
    quicly_stream_t *stream
);
\end{lstlisting}

\hspace{\columnwidth/2 - 0.37cm}\scalebox{1.1}{$\downarrow$} App consumes received bytes
\begin{lstlisting}[linewidth=\columnwidth,breaklines=true]
// Tells quicly how much bytes have been
// consumed from the internal stream buffer.
void quicly_streambuf_ingress_shift(
    quicly_stream_t *stream, size_t delta
);
\end{lstlisting}

\hspace{\columnwidth/2 - 0.48cm}\scalebox{1.7}{$\circlearrowleft$} Back to quicly\_receive()
\end{small}

\texttt{quicly} has this idea to \textit{own} the output buffer and
manage it on behalf of the application under the impulse of the provided API. This
design choice is the cornerstone behind their ability to reassemble QUIC data
fragments and take ownership at the same time. It turns out that in practice,
to take advantage of VReverso, similar intelligence is needed: owning the output
stream buffer is an implementation requirement for the QUIC library, and
therefore the API requires having a call to tell the module how many bytes have
been consumed from the internal buffer. \texttt{quicly} in particular may not
require changing its API to take advantage of VReverso. It is an exception among
the various stacks. Note that, for completeness, it is possible to use
\texttt{quicly}'s low-level API to manipulate independent data from Stream frames
in zero-copy, which may be useful, for example, to deal with HTTP/3's state machine. 
However, quicly cannot buffer QUIC stream data as-is nor handle out-of-order
delivery without copying, due to the QUIC V1 protocol design.  For achieving
\emph{contiguous} zero-copy, we would need \texttt{quicly} to implement QUIC
VReverso.

%% file: reverso.bib
@inproceedings{10.1145/3098822.3098842,
author = {Langley, Adam and Riddoch, Alistair and Wilk, Alyssa and Vicente, Antonio and Krasic, Charles and Zhang, Dan and Yang, Fan and Kouranov, Fedor and Swett, Ian and Iyengar, Janardhan and Bailey, Jeff and Dorfman, Jeremy and Roskind, Jim and Kulik, Joanna and Westin, Patrik and Tenneti, Raman and Shade, Robbie and Hamilton, Ryan and Vasiliev, Victor and Chang, Wan-Teh and Shi, Zhongyi},
title = {{The QUIC Transport Protocol: Design and Internet-Scale Deployment}},
year = {2017},
isbn = {9781450346535},
publisher = {Association for Computing Machinery},
address = {New York, NY, USA},
url = {https://doi.org/10.1145/3098822.3098842},
doi = {10.1145/3098822.3098842},
booktitle = {Proceedings of the Conference of the ACM Special Interest Group on Data Communication},
pages = {183–196},
numpages = {14},
location = {Los Angeles, CA, USA},
series = {SIGCOMM '17}
}

@proceedings{47584,
title	= {{ThinLTO: Scalable and incremental LTO}},
editor	= {Teresa Johnson and Mehdi Amini and Xinliang David Li},
year	= {2017},
booktitle	= {Proceedings of the 2017 International Symposium on Code Generation and Optimization},
pages	= {111--121}
}

@article{pauly2018unreliable,
  title={{An unreliable datagram extension to QUIC}},
  author={Pauly, Tommy and Kinnear, Eric and Schinazi, David},
  journal={Internet Engineering Task Force.(September 2018). draft-pauly-quicdatagram-00},
  year={2018}
}

@inproceedings{de2017multipath,
  title={{Multipath QUIC: Design and evaluation}},
  author={De Coninck, Quentin and Bonaventure, Olivier},
  booktitle={Proceedings of the 13th international conference on emerging networking experiments and technologies},
  pages={160--166},
  year={2017}
}

@incollection{rfc9001,
  title={{RFC 9001: Using TLS to Secure QUIC}},
  author={Thomson, M and Turner, S},
  year={2021},
  publisher={RFC Editor}
}

@misc{rfc5116,
  title={An Interface and Algorithms for Authenticated Encryption},
  author={McGrew, D.},
  publisher={RFC Editor},
  year={2008}
}

@misc{relay-cell,
  title={Relay Cells},
  author={The Tor Project},
  howpublished={\url{https://spec.torproject.org/tor-spec/relay-cells.html}},
  note={Accessed: May 2025}
}

@incollection{rfc9000,
  title={{QUIC: A UDP-based multiplexed and secure transport}},
  author={Iyengar, Jana and Thomson, Martin and others},
  booktitle={RFC 9000},
  year={2021}
}

@misc{rfc9114,
  title={{RFC 9114: HTTP/3}},
  author={Bishop, M},
  year={2022},
  publisher={RFC Editor}
}

@misc{criterion,
  title={{Criterion.rs: Statistics-driven benchmarking library for Rust}},
  author={Brook Heisler},
  howpublished={\url{https://github.com/bheisler/criterion.rs}},
  note={Accessed: June 2023}
}

@misc{wolfssl,
  title={{WolfSSL - Embedded TLS library}},
  author={wolfSSL Inc.},
  howpublished = {\url{https://github.com/wolfSSL/wolfssl}},
  note={Accessed: May 2024},
}

@misc{rustls,
  title={{rustls: A modern TLS library in Rust}},
  author={Birr-Pixton, Joe and Ochtman, Dirkjan and McCarney, Daniel and Aas, Josh},
  howpublished = {\url{https://github.com/rustls/rustls}},
  note={Accessed: May 2024},
}

@misc{gnutls,
  title={{The GnuTLS Transport Layer Security Library}},
  author={Rühsen, Tim and Ueno, Daiki and Baryshkov, Dmitry and Fridrich, Zoltán},
  howpublished={\url{https://www.gnutls.org}},
  note={Accessed: May 2024},
}

@inproceedings{tcpls-hotnets,
author = {Rochet, Florentin and Assogba, Emery and Bonaventure, Olivier},
title = {{TCPLS: Closely Integrating TCP and TLS}},
year = {2020},
isbn = {9781450381451},
publisher = {Association for Computing Machinery},
address = {New York, NY, USA},
url = {https://doi.org/10.1145/3422604.3425947},
doi = {10.1145/3422604.3425947},
booktitle = {Proceedings of the 19th ACM Workshop on Hot Topics in Networks},
pages = {45–52},
numpages = {8},
location = {Virtual Event, USA},
series = {HotNets '20}
}

@inproceedings{tcpls-conext,
  author = {Rochet, Florentin and Assogba, Emery and Piraux, Maxime and Edeline, Korian and Donnet, Benoit and Bonaventure, Olivier},
  title = {{TCPLS: Modern Transport Services with TCP and TLS}},
  year = {2021},
  isbn = {9781450390989},
  publisher = {Association for Computing Machinery},
  address = {New York, NY, USA},
  url = {https://doi.org/10.1145/3485983.3494865},
  doi = {10.1145/3485983.3494865},
  booktitle = {Proceedings of the 17th International Conference on Emerging Networking EXperiments and Technologies},
  pages = {45–59},
  numpages = {15},
  location = {Virtual Event, Germany},
  series = {CoNEXT '21}
}

@article{kohler2006designing,
title="Designing {DCCP}: Congestion control without reliability",
author={Kohler, E. and Handley, M. and Floyd, S.},
journal="{ACM} {SIGCOMM} Computer Communication Review",
volume="36",
number="4",
pages="27--38",
year="2006",
month="August",
}

@techreport{quic-multipath,
	author = {Yanmei Liu and Yunfei Ma and Quentin De Coninck and Olivier 
	Bonaventure and Christian Huitema and Mirja Kuehlewind},
	title = {{Multipath Extension for QUIC}},
	howpublished = {Working Draft},
	type = {Internet-Draft},
	number = {draft-lmbdhk-quic-multipath-00},
	year = {2021},
	month = {October},
	institution = {IETF Secretariat},
	url = {https://www.ietf.org/archive/id/draft-lmbdhk-quic-multipath-00.txt},
}

@misc{stewart2022rfc,
  title={{RFC 9260: Stream Control Transmission Protocol}},
  author={Stewart, Randall and T{\"u}xen, M and Nielsen, K},
  year={2022},
  publisher={RFC Editor}
}

@article{stewart2001sctp,
  title={{SCTP: new transport protocol for TCP/IP}},
  author={Stewart, Randall and Metz, Christopher},
  journal={IEEE Internet Computing},
  volume={5},
  number={6},
  pages={64--69},
  year={2001},
  publisher={IEEE}
}

@inproceedings{tor,
author = {Roger Dingledine and Nick Mathewson and Paul Syverson},
title = {{Tor: The Second-Generation Onion Router}},
booktitle = {13th USENIX Security Symposium (USENIX Security 04)},
year = {2004},
address = {San Diego, CA},
url = {https://www.usenix.org/conference/13th-usenix-security-symposium/tor-second-generation-onion-router},
publisher = {USENIX Association},
month = aug,
}

@article{akdemir2010breakthrough,
  title={{Breakthrough AES performance with intel AES new instructions}},
  author={Akdemir, Kahraman and Dixon, Martin and Feghali, Wajdi and Fay, Patrick and Gopal, Vinodh and Guilford, Jim and Ozturk, Erdinc and Wolrich, Gil and Zohar, Ronen},
  journal={White paper, June},
  volume={11},
  year={2010}
}

@INPROCEEDINGS{detal2013revealing,
author = "Detal, G. and Hesmans, b. and Bonaventure, O. and Vanaubel, Y. and Donnet, B.",
title = "Revealing Middlebox Interference with Tracebox",
booktitle = "Proc. {ACM} Internet Measurement Conference ({IMC})",
year = "2013",
month = "October",
}

@INPROCEEDINGS{hesmans2013tcp,
author = "Hesmans, B. and Duchene, F. and Paasch, C. and Detal, G. and Bonaventure, O.",
title = "Are {TCP} Extensions Middlebox-Proof?",
booktitle = "Proc. Workshop on Hot Topics in Middleboxes and Network Function Virtualization (HotMiddlebox)",
year = "2013",
month = "December",
}

@misc{quiche,
  title = {{quiche: Savoury implementation of the QUIC transport protocol}},
  author = {Ghedini, Alessandro and Pardue, Lucas},
  howpublished={\url{https://github.com/cloudflare/quiche}},
  note={Forked for this paper from commit 8dfc6e0},
}

@misc{quicly,
  title = {{quicly: A modular QUIC stack designed primarily for H2O}},
  author = {Kazuho Oku},
  howpublished = {\url{https://github.com/h2o/quicly}},
  note={Accessed: April 2024}
}

@misc{quinn,
  title = {{quinn: Async-friendly QUIC implementation in Rust}},
  author = {Ochtman, Dirkjan and Saunders, Benjamin and Begue, Jean-Christophe},
  howpublished = {\url{https://github.com/quinn-rs/quinn}},
  note={Accessed: April 2024}
}

@misc{msquic,
  title = {{MsQuic: Cross-platform, C implementation of the IETF QUIC protocol, exposed to C, C++, C\# and Rust.}},
  author = {Nick Banks},
  howpublished = {\url{https://github.com/microsoft/msquic}},
  note={Accessed: April 2024}
}

@misc{picoquic,
  title = {{picoquic: Minimal implementation of the QUIC protocol}},
  author = {Christian Huitema},
  howpublished={\url{https://github.com/private-octopus/picoquic}},
  note={Accessed: April 2024},
}

@misc{xquic,
  title = {{xquic: cross-platform implementation of QUIC and HTTP/3 protocol.}},
  author = {Alibaba},
  howpublished={\url{}https://github.com/alibaba/xquic},
  note={Accessed: April 2024},
}

@misc{rfc9221,
  title={RFC 9221 An Unreliable Datagram Extension to QUIC},
  author={Pauly, T and Kinnear, E and Schinazi, D},
  year={2022},
  type={RFC},
}

@inproceedings{zhou2004reordering,
  title={Reordering of {IP} packets in Internet},
  author={Zhou, Xiaoming and Van Mieghem, Piet},
  booktitle={International workshop on passive and active network measurement},
  pages={237--246},
  year={2004},
  organization={Springer}
}

@misc{roskind2013quic,
  title="{QUIC}, Quick {UDP} Internet Connections",
  author="Roskind, J.",
  howpublished="\url{https://docs.google.com/document/d/1RNHkx_VvKWyWg6Lr8SZ-saqsQx7rFV-ev2jRFUoVD34/preview}",
  year="2013",
  note={Accessed: April 2024}
}

@misc{quic-list,
  title="Implementations list of the {QUIC} protocol",
  author="QUIC WG",
  howpublished="\url{https://github.com/quicwg/base-drafts/wiki/Implementations}",
  note={Accessed: April 2024}
}

@article{papastergiou2016ossifying,
title="De-ossifying the {I}nternet transport layer: A survey and future perspectives",
author="Papastergiou, G. and Fairhurst, G. and Ros, D. and Brunstrom, A. and Grinnemo, K.-J. and Hurtig, P. and Khademi, N. and T{\"u}xen, M. and Welzl, M. and Damjanovic, D. and Mangiante, S.",
journal="{IEEE} Communications Surveys \& Tutorials",
volume="19",
number="1",
pages="619--639",
year="2016",
}

@inproceedings{nowlan2012fitting,
  title="Fitting Square Pegs Through Round Pipes: Unordered Delivery Wire-Compatible with {TCP} and {TLS}",
  author="Nowlan, M. F. and Tiwari, N. and Iyengar, J. and Amin, S. O. and Ford, B.",
  booktitle="Proc. {USENIX} Symposium on Networked Systems Design and Implementation ({NSDI})",
  year="2012",
  month = "April",
}

@inproceedings{jadin2017securing,
title="Securing Multipath {TCP}: Design \& implementation",
author="Jadin, M. and Tihon, G. and Pereira, O. and Bonaventure, O.",
booktitle= "Proc. {IEEE} {INFOCOM}",
year=2017,
month = "May",
}

@misc{michel-thesis,
  title="{Flexible QUIC loss recovery mechanisms for latency-sensitive applications}",
  author="François Michel",
  year ="2023",
  note="PhD. Thesis, UCLouvain"
}

@inproceedings {285094,
  author = {Mariano Scazzariello and Tommaso Caiazzi and Hamid Ghasemirahni and Tom Barbette and Dejan Kosti{\'c} and Marco Chiesa},
  title = {{A High-Speed Stateful Packet Processing Approach for Tbps Programmable Switches}},
  booktitle = {20th USENIX Symposium on Networked Systems Design and Implementation (NSDI 23)},
  year = {2023},
  isbn = {978-1-939133-33-5},
  address = {Boston, MA},
  pages = {1237--1255},
  url = {https://www.usenix.org/conference/nsdi23/presentation/scazzariello},
  publisher = {USENIX Association},
  month = apr
}

@inproceedings{pquic,
author = {De Coninck, Quentin and Michel, Fran\c{c}ois and Piraux, Maxime and Rochet, Florentin and Given-Wilson, Thomas and Legay, Axel and Pereira, Olivier and Bonaventure, Olivier},
title = {{Pluginizing QUIC}},
year = {2019},
isbn = {9781450359566},
publisher = {Association for Computing Machinery},
address = {New York, NY, USA},
url = {https://doi.org/10.1145/3341302.3342078},
doi = {10.1145/3341302.3342078},
booktitle = {Proceedings of the ACM Special Interest Group on Data Communication},
pages = {59–74},
numpages = {16},
location = {Beijing, China},
series = {SIGCOMM '19}
}

@article{michel2022flec,
  title={{FlEC: Enhancing QUIC with application-tailored reliability mechanisms}},
  author={Michel, Fran{\c{c}}ois and Cohen, Alejandro and Malak, Derya and De Coninck, Quentin and M{\'e}dard, Muriel and Bonaventure, Olivier},
  journal={IEEE/ACM Transactions on Networking},
  year={2022},
  publisher={IEEE}
}

@ARTICLE{9433515,
  author={Kosek, Mike and Shreedhar, Tanya and Bajpai, Vaibhav},
  journal={IEEE Communications Magazine},
  title={{Beyond QUIC v1: A First Look at Recent Transport Layer IETF Standardization Efforts}},
  year={2021},
  volume={59},
  number={4},
  pages={24-29},
  doi={10.1109/MCOM.001.2000877}
}

@inproceedings{kuhlewind2021evaluation,
  title={{Evaluation of QUIC-based MASQUE proxying}},
  author={K{\"u}hlewind, Mirja and Carlander-Reuterfelt, Matias and Ihlar, Marcus and Westerlund, Magnus},
  booktitle={Proceedings of the 2021 Workshop on Evolution, Performance and Interoperability of QUIC},
  pages={29--34},
  year={2021}
}

@inproceedings{rogaway2002authenticated,
  title={{Authenticated-Encryption with Associated-Data}},
  author={Rogaway, Phillip},
  booktitle={Proceedings of the 9th ACM Conference on Computer and Communications Security},
  pages={98--107},
  year={2002}
}

@misc{google-datagramchoice,
  title={How the QUIC protocol finds the maximum MTU for a session ?},
  author={Ryan Hamilton},
  howpublished={https://groups.google.com/a/chromium.org/g/proto-quic/c/uKWLRh9JPCo?pli=1},
  note={Accessed: April 2024},
}

@inproceedings{bernstein2012security,
  title={The security impact of a new cryptographic library},
  author={Bernstein, Daniel J and Lange, Tanja and Schwabe, Peter},
  booktitle={Progress in Cryptology--LATINCRYPT 2012: 2nd International Conference on Cryptology and Information Security in Latin America, Santiago, Chile, October 7-10, 2012. Proceedings 2},
  pages={159--176},
  year={2012},
  organization={Springer}
}

@inproceedings{fischlin2015data,
  title={Data is a stream: Security of stream-based channels},
  author={Fischlin, Marc and G{\"u}nther, Felix and Marson, Giorgia Azzurra and Paterson, Kenneth G},
  booktitle={Advances in Cryptology--CRYPTO 2015: 35th Annual Cryptology Conference, Santa Barbara, CA, USA, August 16-20, 2015, Proceedings, Part II 35},
  pages={545--564},
  year={2015},
  organization={Springer}
}

@article{bellare2008authenticated,
  title={Authenticated encryption: Relations among notions and analysis of the generic composition paradigm},
  author={Bellare, Mihir and Namprempre, Chanathip},
  journal={Journal of cryptology},
  volume={21},
  number={4},
  pages={469--491},
  year={2008},
  publisher={Springer}
}

@inproceedings{albrecht2009plaintext,
  title={Plaintext recovery attacks against SSH},
  author={Albrecht, Martin R and Paterson, Kenneth G and Watson, Gaven J},
  booktitle={2009 30th IEEE Symposium on Security and Privacy},
  pages={16--26},
  year={2009},
  organization={IEEE}
}

@inproceedings{brumley2011remote,
  title={Remote timing attacks are still practical},
  author={Brumley, Billy Bob and Tuveri, Nicola},
  booktitle={European Symposium on Research in Computer Security},
  pages={355--371},
  year={2011},
  organization={Springer}
}

@inproceedings{bleichenbacher1998chosen,
  title={Chosen ciphertext attacks against protocols based on the RSA encryption standard PKCS\# 1},
  author={Bleichenbacher, Daniel},
  booktitle={Advances in Cryptology—CRYPTO'98: 18th Annual International Cryptology Conference Santa Barbara, California, USA August 23--27, 1998 Proceedings 18},
  pages={1--12},
  year={1998},
  organization={Springer}
}

@inproceedings{vaudenay2002security,
  title={Security flaws induced by CBC padding—applications to SSL, IPSEC, WTLS...},
  author={Vaudenay, Serge},
  booktitle={International Conference on the Theory and Applications of Cryptographic Techniques},
  pages={534--545},
  year={2002},
  organization={Springer}
}

@inproceedings{rizzo2010practical,
  title={Practical padding oracle attacks},
  author={Rizzo, Juliano and Duong, Thai},
  booktitle={4th USENIX Workshop on Offensive Technologies (WOOT 10)},
  year={2010}
}

@misc{industrial-quic-report,
  title={Accelerating UDP packet transmission for QUIC},
  author={Alessandro Ghedini},
  year={2020},
  howpublished={\url{https://blog.cloudflare.com/accelerating-udp-packet-transmission-for-quic/}},
  note={Accessed: August 2024},
}

@inproceedings{10.1145/3589334.3645323,
author = {Zhang, Xumiao and Jin, Shuowei and He, Yi and Hassan, Ahmad and Mao, Z. Morley and Qian, Feng and Zhang, Zhi-Li},
title = {QUIC is not Quick Enough over Fast Internet},
year = {2024},
isbn = {9798400701719},
publisher = {Association for Computing Machinery},
address = {New York, NY, USA},
url = {https://doi.org/10.1145/3589334.3645323},
doi = {10.1145/3589334.3645323},
booktitle = {Proceedings of the ACM Web Conference 2024},
pages = {2713–2722},
numpages = {10},
location = {Singapore, Singapore},
series = {WWW '24}
}

@misc{dpdk,
  title={Data Plane Development Kit ({DPDK})},
  author={Linux Foundation},
  howpublished={\url{https://www.dpdk.org/}},
  note={Accessed: August 2024}
}

@inproceedings{10.5555/2342821.2342830,
author = {Rizzo, Luigi},
title = {Netmap: a novel framework for fast packet I/O},
year = {2012},
publisher = {USENIX Association},
address = {USA},
booktitle = {Proceedings of the 2012 USENIX Conference on Annual Technical Conference},
pages = {9},
numpages = {1},
location = {Boston, MA},
series = {USENIX ATC'12}
}

@inproceedings {196184,
author = {Kenichi Yasukata and Michio Honda and Douglas Santry and Lars Eggert},
title = {{StackMap}: {Low-Latency} Networking with the {OS} Stack and Dedicated {NICs}},
booktitle = {2016 USENIX Annual Technical Conference (USENIX ATC 16)},
year = {2016},
isbn = {978-1-931971-30-0},
address = {Denver, CO},
pages = {43--56},
url = {https://www.usenix.org/conference/atc16/technical-sessions/presentation/yasukata},
publisher = {USENIX Association},
month = jun
}

@inproceedings{eggert2020towards,
  title={Towards securing the Internet of Things with QUIC},
  author={Eggert, Lars},
  booktitle={Proc. Workshop Decentralized IoT Syst. Security (DISS)},
  pages={1--6},
  year={2020}
}

@inproceedings{marx2020same,
  title={Same standards, different decisions: A study of QUIC and HTTP/3 implementation diversity},
  author={Marx, Robin and Herbots, Joris and Lamotte, Wim and Quax, Peter},
  booktitle={Proceedings of the Workshop on the Evolution, Performance, and Interoperability of QUIC},
  pages={14--20},
  year={2020}
}

@inproceedings{180723,
author = {Costin Raiciu and Christoph Paasch and Sebastien Barre and Alan Ford and Michio Honda and Fabien Duchene and Olivier Bonaventure and Mark Handley},
title = {How Hard Can It Be? Designing and Implementing a Deployable Multipath {TCP}},
booktitle = {9th USENIX Symposium on Networked Systems Design and Implementation (NSDI 12)},
year = {2012},
isbn = {978-931971-92-8},
address = {San Jose, CA},
pages = {399--412},
url = {https://www.usenix.org/conference/nsdi12/technical-sessions/presentation/raiciu},
publisher = {USENIX Association},
month = apr
}

@inproceedings{10.1145/1866307.1866363,
author = {Degabriele, Jean Paul and Paterson, Kenneth G.},
title = {{On the (in)security of IPsec in MAC-then-encrypt configurations}},
year = {2010},
isbn = {9781450302456},
publisher = {Association for Computing Machinery},
address = {New York, NY, USA},
url = {https://doi.org/10.1145/1866307.1866363},
doi = {10.1145/1866307.1866363},
booktitle = {Proceedings of the 17th ACM Conference on Computer and Communications Security},
pages = {493–504},
numpages = {12},
location = {Chicago, Illinois, USA},
series = {CCS '10}
}

@InProceedings{10.1007/3-540-46035-7_35,
author="Vaudenay, Serge",
editor="Knudsen, Lars R.",
title={{Security Flaws Induced by CBC Padding --- Applications to SSL, IPSEC, WTLS...}},
booktitle="Advances in Cryptology --- EUROCRYPT 2002",
year="2002",
publisher="Springer Berlin Heidelberg",
address="Berlin, Heidelberg",
pages="534--545",
isbn="978-3-540-46035-0"
}

@InProceedings{10.1007/978-3-540-45146-4_34,
author="Canvel, Brice
and Hiltgen, Alain
and Vaudenay, Serge
and Vuagnoux, Martin",
editor="Boneh, Dan",
title={{Password Interception in a SSL/TLS Channel}},
booktitle="Advances in Cryptology - CRYPTO 2003",
year="2003",
publisher="Springer Berlin Heidelberg",
address="Berlin, Heidelberg",
pages="583--599",
isbn="978-3-540-45146-4"
}

@InProceedings{10.1007/978-3-642-25385-0_20,
author="Paterson, Kenneth G.
and Ristenpart, Thomas
and Shrimpton, Thomas",
editor="Lee, Dong Hoon
and Wang, Xiaoyun",
title={{Tag Size Does Matter: Attacks and Proofs for the TLS Record Protocol}},
booktitle="Advances in Cryptology -- ASIACRYPT 2011",
year="2011",
publisher="Springer Berlin Heidelberg",
address="Berlin, Heidelberg",
pages="372--389",
isbn="978-3-642-25385-0"
}

@inproceedings{Paterson2012PlaintextRecoveryAA,
  title={{Plaintext-Recovery Attacks Against Datagram TLS}},
  author={Kenneth G. Paterson and Nadhem J. AlFardan},
  booktitle={Network and Distributed System Security Symposium},
  year={2012},
}

@INPROCEEDINGS{6547131,
  author={Al Fardan, Nadhem J. and Paterson, Kenneth G.},
  booktitle={2013 IEEE Symposium on Security and Privacy},
  title={{Lucky Thirteen: Breaking the TLS and DTLS Record Protocols}},
  year={2013},
  volume={},
  number={},
  pages={526-540},
  doi={10.1109/SP.2013.42}
}

@InProceedings{10.1007/978-3-662-49890-3_24,
author="Albrecht, Martin R.
and Paterson, Kenneth G.",
editor="Fischlin, Marc
and Coron, Jean-S{\'e}bastien",
title={{Lucky Microseconds: A Timing Attack on Amazon's s2n Implementation of TLS}},
booktitle="Advances in Cryptology -- EUROCRYPT 2016",
year="2016",
publisher="Springer Berlin Heidelberg",
address="Berlin, Heidelberg",
pages="622--643",
isbn="978-3-662-49890-3"
}

@inproceedings{10.1145/3458336.3465283,
author = {Wolnikowski, Adam and Ibanez, Stephen and Stone, Jonathan and Kim, Changhoon and Manohar, Rajit and Soul\'{e}, Robert},
title = {Zerializer: towards zero-copy serialization},
year = {2021},
isbn = {9781450384384},
publisher = {Association for Computing Machinery},
address = {New York, NY, USA},
url = {https://doi.org/10.1145/3458336.3465283},
doi = {10.1145/3458336.3465283},
booktitle = {Proceedings of the Workshop on Hot Topics in Operating Systems},
pages = {206–212},
numpages = {7},
location = {Ann Arbor, Michigan},
series = {HotOS '21}
}

@inproceedings{10.1145/3458336.3465287,
author = {Raghavan, Deepti and Levis, Philip and Zaharia, Matei and Zhang, Irene},
title = {Breakfast of champions: towards zero-copy serialization with NIC scatter-gather},
year = {2021},
isbn = {9781450384384},
publisher = {Association for Computing Machinery},
address = {New York, NY, USA},
url = {https://doi.org/10.1145/3458336.3465287},
doi = {10.1145/3458336.3465287},
booktitle = {Proceedings of the Workshop on Hot Topics in Operating Systems},
pages = {199–205},
numpages = {7},
location = {Ann Arbor, Michigan},
series = {HotOS '21}
}

@inproceedings{10.1145/3600006.3613137,
author = {Raghavan, Deepti and Ravi, Shreya and Yuan, Gina and Thaker, Pratiksha and Srivastava, Sanjari and Murray, Micah and Penna, Pedro Henrique and Ousterhout, Amy and Levis, Philip and Zaharia, Matei and Zhang, Irene},
title = {Cornflakes: Zero-Copy Serialization for Microsecond-Scale Networking},
year = {2023},
isbn = {9798400702297},
publisher = {Association for Computing Machinery},
address = {New York, NY, USA},
url = {https://doi.org/10.1145/3600006.3613137},
doi = {10.1145/3600006.3613137},
booktitle = {Proceedings of the 29th Symposium on Operating Systems Principles},
pages = {200–215},
numpages = {16},
location = {Koblenz, Germany},
series = {SOSP '23}
}

@inproceedings{qian2011profiling,
  title={Profiling resource usage for mobile applications: a cross-layer approach},
  author={Qian, Feng and Wang, Zhaoguang and Gerber, Alexandre and Mao, Zhuoqing and Sen, Subhabrata and Spatscheck, Oliver},
  booktitle={Proceedings of the 9th international conference on Mobile systems, applications, and services},
  pages={321--334},
  year={2011}
}

@article{segovia2023efficiency,
  title={Efficiency traps beyond the climate crisis: exploration--exploitation trade-offs and rebound effects},
  author={Segovia-Martin, Jose and Creutzig, Felix and Winters, James},
  journal={Philosophical Transactions of the Royal Society B},
  volume={378},
  number={1889},
  pages={20220405},
  year={2023},
  publisher={The Royal Society}
}
